\documentclass[%
 reprint,
nofootinbib,
 amsmath,amssymb,
 aps, prd
]{revtex4-1}

\usepackage{graphicx}% Include figure files
\usepackage{dcolumn}% Align table columns on decimal point
\usepackage{bm}% bold math
\usepackage{color}    
\usepackage{slashed} 
\usepackage{scalerel}
\usepackage{url}
\usepackage[pdfpagelabels, pdfencoding=auto, psdextra]{hyperref}
\hypersetup{%
 pdfsubject=Paper,
 unicode = true,
 breaklinks = true,
 colorlinks = true,
 linkcolor = blue,
 citecolor = blue,
 menucolor = blue,
 citecolor = blue,
 urlcolor = blue
}

\begin{document}
\setlength{\tabcolsep}{4pt}
\renewcommand{\arraystretch}{1.4}
\renewcommand{\vec}{\boldsymbol}
\newcommand{\mq}{m_{\mathrm{q}}}
\newcommand{\ms}{m_{\mathrm{s}}}
\newcommand{\ma}{m_{\mathrm{a}}}
\newcommand{\mn}{M_{\mathrm{N}}}
\newcommand{\md}{M_{\mathrm{\Delta}}}
\newcommand{\mpi}{M_\mathrm{\pi}}
\newcommand{\half}{\frac{1}{2}}
\newcommand{\kperp}[1]{\vec{#1}_\perp}
\newcommand{\Fn}[1]{F_{#1N}}
\newcommand{\FAn}[1]{\widetilde{F}_{#1N}}
\newcommand{\Fd}[1]{F_{#1\Delta}}
\newcommand{\FAd}[1]{\widetilde{F}_{#1\Delta}}
\newcommand{\ged}{F_{_{N\Delta}}^{\scaleto{E}{4pt}} }
\newcommand{\gmd}{F_{_{N\Delta}}^{\scaleto{M}{4pt}}}
\renewcommand{\gcd}{F_{_{N\Delta}}^{\scaleto{C}{4pt}}}
\newcommand{\ashged}{G_{_{N\Delta}}^{\scaleto{E}{4pt}} }
\newcommand{\ashgmd}{G_{_{N\Delta}}^{\scaleto{M}{4pt}}}
\newcommand{\ashgcd}{G_{_{N\Delta}}^{\scaleto{C}{4pt}}}
\newcommand{\ca}[1]{C_{#1}^A}
\newcommand{\gddezero}{G^{\scaleto{E}{4pt}\scaleto{0}{4pt}}_{_{\Delta\Delta}}}
\newcommand{\gddetwo}{G^{\scaleto{E}{4pt}\scaleto{2}{4pt}}_{_{\Delta\Delta}}}
\newcommand{\gddmone}{G^{\scaleto{M}{4pt}\scaleto{1}{4pt}}_{_{\Delta\Delta}}}
\newcommand{\gddmthree}{G^{\scaleto{M}{4pt}\scaleto{3}{4pt}}_{_{\Delta\Delta}}}
\newcommand{\mmu}{m_\mu}
\newcommand{\xilin}[1]{\textcolor{red}{#1}}
\newcommand{\II}{({\bf II}) \/}
\newcommand{\III}{({\bf III})\/}

\preprint{NT@UW-19-19}
\preprint{INT-PUB-19-060}

\title{Unified model of nucleon elastic form factors and implications for neutrino-oscillation experiments}% Force line breaks with \\

\author{Xilin Zhang}
\email{zhang.10038@osu.edu}
\affiliation{Department of Physics, The Ohio State University, Columbus, OH 43210, USA }
\affiliation{Department of Physics, University of Washington, Seattle, WA \ \ 98195, USA}%

\author{T.~J.~Hobbs}
\email{tjhobbs@smu.edu}
\affiliation{Department of Physics, Southern Methodist University, Dallas, TX \ \ 75275, USA}
\affiliation{Jefferson Lab, EIC Center, Newport News, VA 23606, USA}
\affiliation{Department of Physics, University of Washington, Seattle, WA \ \ 98195, USA}%

\author{Gerald A. Miller}
\email{miller@uw.edu}
\affiliation{Department of Physics, University of Washington, Seattle, WA \ \ 98195, USA}%

\date{\today}% It is always \today, today,

\begin{abstract}
Precise knowledge of the nucleon's axial-current form factors is crucial for modeling GeV-scale neutrino-nucleus interactions.
Unfortunately, the axial form factor remains insufficiently constrained to meet the precision requirements of upcoming long-baseline neutrino-oscillation experiments. This work studies the nucleon's axial and vector form factors  using the light-front approach to build a quark-diquark model of the nucleon
with an explicit pion cloud. The light-front wave functions in both the quark and pion-baryon Fock spaces are first calibrated to existing experimental information on the  nucleon's electromagnetic form factors, and then used to predict the axial form factor. The resulting squared charge radius of the axial pseudo-vector form factor is predicted to be  $r_A^2\! =\! 0.29\! \pm\! 0.03\, \mathrm{fm}^2$, where the small error accounts for the model's parametric uncertainty. We use our form factor results to explore the (quasi-)elastic scattering of neutrinos by (nuclei)nucleons, with the result that the  the widely-implemented dipole ansatz is  an inadequate approximation of the full form factor for modeling both processes. The approximation leads to a $5\!-\!10\%$ over-estimation of the total cross section, depending on the (anti)neutrino energy. We project over-estimations of similar size in the flux-averaged cross sections for the upcoming DUNE long-baseline neutrino-oscillation experiment. 
\end{abstract}

\pacs{Valid PACS appear here}% PACS, the Physics and Astronomy
\maketitle

%\tableofcontents

\section{\label{sec:intro} Introduction}
Modern investigations along the Intensity Frontier~\cite{P5} aim to test the Standard Model (SM) and explore the origins of neutrino mass through a dedicated series of neutrino-oscillation searches,
which rely on the scattering of high-intensity neutrino beams by nuclear targets. At the present time, the dominant limitations in these experiments are an imperfect determination of the
the neutrino flux, and imprecision in theoretical predictions for neutrino-nucleus cross sections, both of which are necessary to extract the neutrino (dis)appearance rates between
the near- and far-detectors in long-baseline measurements.
Improving the theoretical description of neutrino-nucleus reactions in the multiple-GeV neutrino-energy region is therefore critical for the next-generation long-baseline neutrino-oscillation
experiments~\cite{Alvarez-Ruso:2017oui}. In most theoretical frameworks~\cite{Alvarez-Ruso:2017oui}, the neutrino-nucleon interaction is the most basic input to the calculation, such
that the neutrino-nucleon scattering/reaction is the fundamental kernel. As such, the nucleon-level kernels must be carefully investigated in order to understand their accuracy and
potential model uncertainties, as well as to the resulting implications for calculations of nucleus-level scatterings/reactions. Such an understanding can then provide guidance for
further improvements. In those regions of the neutrino energy ($E_\nu$) for which the neutrino-nucleus cross section is dominated by quasi-elastic (QE) scattering and resonance
production~\cite{Formaggio:2013kya}, the nucleon-level kernel requires detailed knowledge of the (in)elastic nucleon form factors of the electroweak (EW) current, including
the axial-current component~\cite{Bernard:2001rs} (the axial form factor). Unfortunately, the axial-current component of the EW form factors remains insufficiently understood
to meet the precision objectives of the coming neutrino-oscillation experiments~\cite{Alvarez-Ruso:2017oui,Hill:2017wgb}. 

In principle, Lattice QCD calculations could provide reliable results about these EW form
factors~\cite{Alvarez-Ruso:2017oui,Kronfeld:2019nfb,Green:2017keo,Rajan:2017lxk,Capitani:2017qpc,Jang:2018lup,Ishikawa:2018rew,Bali:2018qus,Shintani:2018ozy,Jang:2018djx}. However,
these calculations are restricted to a finite window of momentum transfer, $Q$ ({\it i.e.}, $Q^2\!\sim\! 1\, \mathrm{GeV}^2$). Beyond this, a systematic description of the higher-$Q^2$,
several-$\mathrm{GeV}^2$ regime---a region in which the form factors are unlikely to achieve their asymptotic $Q^2$ dependence---is still needed. Moreover, Lattice QCD calculations
for the axial form factor remain generally challenging, with the inelastic form factors expected to be all the more so.

Other currently available frameworks are mainly composed of phenomenological fits of data such as polynomial-based fits (see, {\it e.g.}, Ref.~\cite{Kelly:2004hm}), the
$z$-expansion method,  which entails minimal model dependence~\cite{Bhattacharya:2011ah,Meyer:2016oeg, Hill:2017wgb}, 
  quark-hadron-duality constrained fits~\cite{Bodek:2007ym}, and recent neural-network based fits~\cite{Alvarez-Ruso:2018rdx}.  
 In addition, there are dispersion analyses mixed with the
meson-dominance picture~\cite{Perdrisat:2006hj,Pacetti:2015iqa}, effective field theory approaches focused on the low-$Q^2$ region~\cite{Bernard:2001rs,Bernard:1998gv,Schindler:2006it,Schindler:2006jq,Ando:2006xy,Perdrisat:2006hj,Scherer:2009bt,Yao:2017fym},
and various quark models~\cite{Perdrisat:2006hj}. In this work, we start with the last approach, in particular, the light-front quark
model~\cite{Chung:1991st,Cardarelli:1995dc,Brodsky:1997de, Miller:2002ig, Miller:2002qb,Ma:2002ir,Ma:2002xu,Pasquini:2007iz,Cloet:2012cy,Punjabi:2015bba,Brodsky:2014yha}. It is well
known that pionic degrees-of-freedom are important aspect of the dynamics of the strong-interaction, being responsible for the long-distance structure of the nucleon's charge structure.
For this reason, we manifestly include 
contributions from the nucleon's pion cloud~\cite{Thomas:1981vc,Miller:2002ig, Miller:2002qb,Pasquini:2007iz,Cloet:2012cy} in our model.
With this approach the nucleon's wave function is  governed  by a mixture of contributions from a quark-diquark core and pion cloud, the latter due to the reconfiguration of the nucleon
into pion-baryon intermediate modes [see Eq.~(\ref{eqn:nwfdecomp1})]. 

In contrast to the other non-lattice approaches we noted, our model is capable of simultaneously describing 
the elastic electromagnetic (EM) and axial form factors  in the $ Q^2\!\sim\!\mathrm{few\!-\!GeV}^2$ range. 
 (The framework can also be generalized to study the inelastic form factors.)
It thus unifies  these various form factors in a single approach, which is valuable considering the
large amount of experimental information for the EM   elastic form factors, which might be exploited to improve the axial form factors. To realize and demonstrate these connections, our model, including the quark's light-front wave function, is first calibrated against the better-determined nucleon elastic EM form factors, and then used to predict the elastic axial  form factors. 

By evaluating the first derivative with respect to $Q^2$ of the axial pseudo-vector form factor [$\FAn{1}$, see the definition in Eq.~(\ref{eqn:Axialmatrixelement})], {\it i.e.},
$r_A^2 \equiv -\frac{6}{\FAn{1}} \frac{d \FAn{1}}{d Q^2}\vert_{Q^2=0} $, we obtain the nucleon's axial-charge radius, $r_A^2\! =\! 0.29\! \pm\! 0.03\, \mathrm{fm}^2$, which should be compared
to $r_A^2\! =\! 0.46\! \pm\! 0.16 \mathrm{fm}^2$ from a combined analysis~\cite{Hill:2017wgb} of neutrino-nucleon scattering data and the singlet muonic hydrogen capture-rate measurement; and
also to current Lattice QCD results, which range from $r_A^2\! =\! 0.2$ to $0.45\, \mathrm{fm}^2$. If we match our form factor and its derivative to a dipole parameterize,
$g_A \tilde{G}_D(Q^2) \equiv g_A/(1+ Q^2/M_A^2)^{2}$, at $Q^2=0$, the single mass-parameter, $M_A$, is then given as $\sqrt{12/r_A^2}$, and for it we predict
$M_A\! =\! 1.28\! \pm\! 0.07$ GeV. 

We stress, however, that such an approximation would seriously over-estimate the (anti)neutrino-nucleon cross sections compared to calculations based on the full expression of the form factor,
by 5-10\% for $E_\nu\! \gtrsim\! 0.5$ GeV. Consequently, fitting the dipole approximation to the full form factor over a range of $Q^2$ ($M_A$ is then not related to $r_A$), would be
expected to produce an effective $M_A$ smaller than $\sim 1.28$ GeV. Nevertheless, it will still be larger than the central value of the recent analysis:  $1.01 \pm 0.17\, \mathrm{GeV}^2$,
based on their $r_A^2$ results~\cite{Meyer:2016oeg,Hill:2017wgb}, since the (anti)neutrino-nucleon cross section given by the full form factor is intermediate between the results using the
two dipole approximations with $M_A=1$ and $1.28$ GeV (see Fig.~\ref{fig:sig}). 

To further assess how these discrepancies with the dipole approximation can be expected to impact neutrino cross sections, we implement the axial form factors in a simulation of
neutrino-$^{40}\rm{Ar}$ QE scattering using the GiBUU event generator~\cite{Buss:2011mx}, and compute the flux-averaged cross sections based on the energy distribution of the projected
neutrino flux at DUNE~\cite{Acciarri:2015uup}. Here, we again find that the discrepancy leads to $5\%$ overestimate of the cross sections for both neutrino and antineutrino scatterings
at $Q^2\! <\! 0.2\, \mathrm{GeV}^2$---the peak location of the flux-averaged differential cross section, $d\sigma/dQ^2$---and climb to $10\!-\!15\%$ at larger $Q^2$
(see Fig.~\ref{fig:dsigdQsqLBNE}). Meanwhile, the over-estimation of the neutrino and antineutrino scattering cross section is similar at $Q^2\! <\! 0.5\, \mathrm{GeV}^2$, but still
differs at the few-percent level at larger $Q^2$. 

In the remainder of this article, we detail in Sec.~\ref{sec:formalism} the theory formalism for our pion-cloud-augmented light-front quark model. Sec.~\ref{sec:input} discusses the input
parameters for the model, while Sec.~\ref{sec:constraints} presents our procedure for constraining the unknown parameters in the model via measurements of the nucleon's EM form factors,
and the resulting predictions for the axial-current form factor $\FAn{1}$. In Sec.~\ref{sec:Impacts}, we first discuss these form factors' impacts on the single-nucleon cross sections,
and then their impacts on the flux-averaged cross sections for neutrino-${}^{40}\rm{Ar}$ QE scattering. A short summary with conclusions is provided in Sec.~\ref{sec:conc}. Readers interested
mainly in the final analysis for neutrino-nucleus scattering can directly consult Sec.~\ref{sec:Impacts} and possibly Sec.~\ref{sec:constraints}, which demonstrate the success of our
model in reproducing the EM form factors. Explanations of relevant notation can be found in Sec.~\ref{sec:formalism}. 

\section{\label{sec:formalism} Formalism } 

\subsection{ The model }
 
The nucleon's wave function in the framework of the light-front quark model~\cite{Brodsky:1997de, Miller:2002ig, Miller:2002qb,Pasquini:2007iz,Cloet:2012cy} can be schematically written as
\begin{eqnarray}
\vert p_{_N}, \lambda_{_N}; N \rangle\, &&= \sqrt{Z} \vert p_{_N}, \lambda_{_N}; N \rangle_\mathrm{q\otimes d}   \notag \\
	                              &&+\, \vert p_{_N}, \lambda_{_N}; N \rangle_\mathrm{B\otimes \pi} \ , \label{eqn:nwfdecomp1}
\end{eqnarray}
with the first component being in terms of quark-diquark ($\mathrm{q\otimes d} $) degrees-of-freedom, and the second in terms of hadronic ({\it i.e.}, baryon and pion, $\mathrm{B\otimes \pi}) $ degrees-of-freedom. In this work we simplify the
quark-level description of the nucleon as consisting of a quark and a two-body quark$\oplus$quark spectator, known as a diquark~\cite{Cloet:2012cy}.  The second component of Eq.~(\ref{eqn:nwfdecomp1}) accounts for contributions from the pion cloud, which is known to accompany the nucleon and $\Delta$ resonances~\cite{Miller:2002ig, Miller:2002qb,Pasquini:2007iz,Cloet:2012cy}.
These two components are orthogonal, {\it i.e.}, $_{B\otimes \pi}\langle p_{_N}, \lambda_{_N}; N \vert p_{_N}, \lambda_{_N}; N \rangle_\mathrm{q\otimes d} = 0$.

The nucleon's EW current form factors can be extracted from the corresponding EM and axial current matrix elements,  
\begin{eqnarray}
&&\langle p_{_{N}}' \lambda_{_N}'; N\vert J^\mu_\mathrm{EM}\left(0\right)   \vert p_{_N }, \lambda_{_N}; N \rangle  \notag \\ 
\quad &&\equiv \bar{u}(p_{_N}',\lambda_{_N}')\left[\Fn{1} \gamma^\mu + \Fn{2} \frac{i\sigma^{\mu\nu} q_\nu}{2 \mn}   \right]  u(p_{_N },\lambda_{_N}) \ ,   \label{eqn:EMmatrixelement}
\end{eqnarray} 
and 
\begin{eqnarray}
&&\langle p_{_{N}}' \lambda_{_N}'; N\vert \vec{J}^\mu_A\left(0\right)   \vert p_{_N }, \lambda_{_N}; N \rangle \notag  \\ 
\quad &&\equiv \bar{u}(p_{_N}',\lambda_{_N}')\left[\FAn{1} \gamma^\mu \gamma_5  + \FAn{2} \frac{q^{\mu}\gamma_5}{2 \mn}   \right] \frac{\vec{\tau}}{2}  u(p_{_N },\lambda_{_N})  \ . 
\label{eqn:Axialmatrixelement}
\end{eqnarray}
Here, the momentum transfer is $q^\nu \equiv \left(p_{_N}'-p_{_N}\right)^\nu$ with $Q^2\equiv - q^\nu q_\nu $, and $N$ denotes either a proton or neutron. The form factors $F_1$, $F_2$,
$\FAn{1}$, and $\FAn{2}$ are all functions of $Q^2$. We also note that the axial current, $\vec{J}^\mu_A$, is a vector in isospin space.  In the following, we especially focus on $\FAn{1}$, while
$\FAn{2}$ can be related to $\FAn{1}$ via the Goldberger-Treiman relation~\cite{Bernard:2001rs}.

Relying on the methods of light-front quantization \cite{Brodsky:1997de, Cloet:2012cy}, the form factors can be extracted from the matrix elements of
Eqs.~(\ref{eqn:EMmatrixelement})-(\ref{eqn:Axialmatrixelement}) by simply studying the plus-components of the currents as
\begin{align}
F_{1 N } &=  \frac{1}{2  p_{_{N}}^+} \langle p_{_{N}}', \lambda_{_N}'=\half; N\vert J^+_\mathrm{EM}\vert p_{_N }, \lambda_{_N}=\half; N \rangle \ , \label{eqn:F1N}  \\ 
F_{2 N } &=  -\frac{\sqrt{2}\mn}{q^R} \frac{1}{2  p_{_{N}}^+}  \label{eqn:F2N} \\
	&\times \langle p_{_{N}}', \lambda_{_N}'=-\half; N\vert J^+_\mathrm{EM}\vert p_{_N }, \lambda_{_N}=\half; N \rangle  \notag \\ 
\FAn{1} & \langle N \vert\frac{\vec{\tau}}{2}\vert N \rangle = \frac{1}{2  p_{_{N}}^+} \label{eqn:GAN} \\
	&\times \langle p_{_{N}}', \lambda_{_N}'=\half; N\vert \vec{J}^+_A \vert p_{_N }, \lambda_{_N}=\half; N \rangle \ . \notag
\end{align}
In the light-front quantization, the time and longitudinal components of 4-vectors (such as current and momentum) are now transformed to the $\pm$ components, {\it e.g.}, for $ q^\mu$, $q^\pm \equiv  q^0 \pm q^z$~\cite{Brodsky:1997de}; for the transverse components, a specific index notation is introduced~\cite{Pasquini:2007iz}: {\it e.g.}, $q^R \equiv -(q^x+iq^y)/\sqrt{2}$ and $q^L \equiv (q^x - iq^y)/\sqrt{2}$.

We point out that other combinations of initial/final nucleon helicities are trivially related to those given in the Eqs.~above~\cite{Pasquini:2007iz,Cloet:2012cy}. 

\begin{figure}
\includegraphics[width=2cm, angle=0]{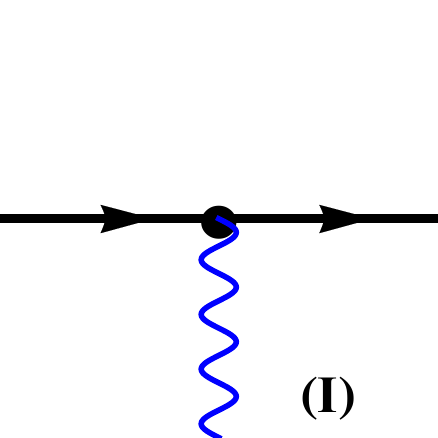} \quad\quad 
\includegraphics[width=2cm, angle=0]{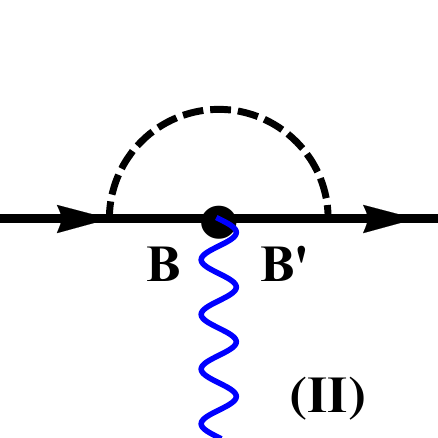}  \quad \quad 
\includegraphics[width=2cm, angle=0]{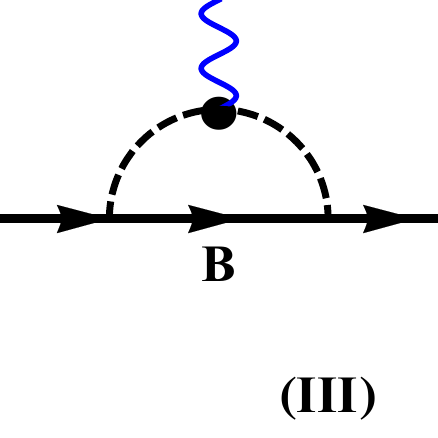}
	\caption{\label{fig:FeynmannD} The diagrammatic representation of the form factor calculations. In ({\bf II}) and ({\bf III}), $\mathrm{B}$ and $\mathrm{B}'$ represent baryon, which can be N or
	$\Delta$.  In ({\bf II}), $\mathrm{B}$ and $\mathrm{B}'$ can be different, meaning that both inelastic and elastic form factors can contribute here.}
\end{figure}

On the basis of the wave-function decomposition in Eq.~(\ref{eqn:nwfdecomp1}), the form factor calculations---equivalent to the above matrix-element calculations---can be represented in terms of the
diagrams shown in Fig.~\ref{fig:FeynmannD}, each of which represents a distinct contribution to the form factor model. Diagram ({\bf I}) represents the contributions from the bare the quark-diquark
configuration terms in Eq.~(\ref{eqn:nwfdecomp1}), while Diagrams ({\bf II}) and ({\bf III}) are from the other Fock space components, in which the nucleon dissociates into pion-baryon states. The
external EW probe is allowed to couple to either the intermediate baryon [in Diagram ({\bf II})] or the recoiling pion [in Diagram ({\bf III})], and both processes contribute to the full model. A possible additional graph involving the direct coupling of the external boson to the $\pi N$ vertex is effectively included when the pseudoscalar pion-nucleon coupling
[see Eq.~(\ref{eq:lagrange}) below] is used. This is because the isovector combination of the $\gamma N \to \pi N$ Born terms that is included in our calculation reproduces the direct $\gamma\pi N$ coupling.

The current set of interactions is consistent with the partially conserved axial vector current within our approximation
scheme~\cite{Miller:2002ig,Miller:2002qb,Cloet:2012cy,Morgan:1985kr}. An additional term involving a direct $aN\pi$ coupling, with $a$
denoting an external axial source, may also contribute~\cite{Hill:2019xqk}. Previous
experience~\cite{Miller:2002ig,Miller:2002qb,Cloet:2012cy,Morgan:1985kr} indicates that possible effects of such direct terms are
approximately accounted for within the parameter variations to be discussed below.

In the following subsections, we proceed in order, relying on the Diagrams ({\bf I})-({\bf III}) to compute the required matrix elements in the light-front quantization. Thus, in
Sec.~\ref{subsec:barequark} we first compute the bare quark-diquark contributions contained in Diagram ({\bf I}), and present in Sec.~\ref{subsec:pion} the pion-cloud pieces
from Diagrams ({\bf II}) and ({\bf III}).

\subsection{\label{subsec:barequark} Diagram (I)}
\begin{widetext}

The quark Fock-space wave function has two components each in spin and flavor space, and we therefore use an $\mathrm{SU}(4)$ ansatz to combine these two spaces~\cite{Meyer:1990fr,Cloet:2012cy}.
For instance, for the proton, the spin-flavor wave function is   
\begin{eqnarray}
\vert \lambda_{_P}; P \rangle_\mathrm{q\otimes d} = \frac{1}{\sqrt{2}}\vert \lambda_{_P}; P \rangle^{\mathrm{f.s.}} \vert \lambda_{_P}; P \rangle^{\mathrm{s.s.}}\vert  \mathrm{CM} \rangle  + \frac{1}{\sqrt{2}} \vert \lambda_{_P}; P \rangle^{\mathrm{f.t.}} \vert \lambda_{_P}; P \rangle^{\mathrm{s.t.}}  \vert \mathrm{CM} \rangle     \ . \label{eqn:qdqwfdef1}
\end{eqnarray}
Here, f.s. and f.t. refer to the flavor-singlet and flavor-triplet states of the diquark system, while s.s.~and s.t.~represent its spin-singlet and spin-triplet states, respectively. The degrees-of-freedom identified
with the center-of-mass (CM) motion are manifestly factorized in this definition, such that the other components are associated with the relative motion degrees-of-freedom. 
In the spectator picture, supposing the quark interacting with the current is a $u$-quark, for instance, we then have 
\begin{eqnarray}
\vert \lambda_{_P}; P \rangle^{\mathrm{f.s.}} & = & \frac{1}{\sqrt{2}} \left(\vert u u d \rangle - \vert u d u\rangle \right) \equiv  \vert u \left(ud\right)^s \rangle  \\   
\vert \lambda_{_P}; P \rangle^{\mathrm{f.t.}} & = &  \frac{1}{\sqrt{6}} \left(\vert u u d \rangle  +  \vert u d u \rangle -2 \vert  d u u \rangle \right) \equiv \sqrt{\frac{1}{ 3}} \vert u \left(ud\right)^{t,0} \rangle - \sqrt{\frac{2}{ 3 }} \vert d \left(ud\right)^{t,1}\rangle\, .
\end{eqnarray}
Meanwhile, we assume the wave functions in spin space, ``s.s.'' and  ``s.t.'' are associated with the scalar and axial-vector diquark respectively~\cite{Brodsky:2003pw,Cloet:2012cy}. The diquarks have definite masses, $\ms$ and $\ma$, and furthermore their wave functions are independent of quark flavor. They can be written as  
\begin{eqnarray}
\vert p_{_N}^+, \vec{p}_{_N \perp} , \lambda_{_N}; N \rangle^{s.s} &= & \int \frac{d x d \vec{k}_\perp}{16\pi^3 x (1-x)}  \sum_{\lambda_q} \phi^{\lambda_{_N}}_{\lambda_q }\left(x,\kperp{k}\right)  \vert x,\kperp{k},\lambda_q ; q,d=s\rangle  \, \label{eqn:qdqdef1} \\  
\vert p_{_N}^+, \vec{p}_{_N \perp} , \lambda_{_N}; N \rangle^{s.t} &  =  & \int \frac{d x d \vec{k}_\perp}{16\pi^3 x (1-x)}  \sum_{\lambda_q \lambda_d} \phi^{\lambda_{_N}}_{\lambda_q \lambda_d}\left(x,\kperp{k}\right)  \vert x,\kperp{k},\lambda_q,\lambda_d ; q,d=a\rangle \ . \label{eqn:qdqdef2} 
\end{eqnarray}
Note the normalization of a {\it single}-particle state is $\langle p^{+'} \kperp{p}'  \vert p^+,\kperp{p} \rangle = (2\pi)^3 2 p^+ \delta(p^+-p^{+'})\delta(\kperp{p}-\kperp{p})$, so the two-particle state's normalization can be written in a fashion with the CM motion manifestly factorized out: $\langle p_1', p_2'  \vert p_1, p_2 \rangle =(2\pi)^3 2P^+\delta(P^+-P^{+'})\delta(\kperp{P}-\kperp{P}') (2 \pi)^3 2 x (1-x)\delta(x-x') \delta(\kperp{k}-\kperp{k}')$. Since the CM is already factorized out in Eq.~(\ref{eqn:qdqwfdef1}), the normalization of the quark-Fock-space basis  for relative motion is $\langle x',\kperp{k}',\lambda_{q}',\lambda_{d}' ; q,d   \vert x,\kperp{k},\lambda_q,\lambda_d ; q,d\rangle  =16\pi^3  x(1-x) \delta^{\lambda_q}_{\lambda_{q}'} \delta^{\lambda_d}_{\lambda_{d}'} \delta\left(x-x'\right)\delta\left(\kperp{k}-\kperp{k}'\right)$. Moreover, the convention for kinematic variables is that the struck quark carries momentum fraction $x$, and transverse momentum $\kperp{k}$, with the spectator having $1-x$ and $-\kperp{k}$ in the CM frame. The intrinsic wave function, {\it e.g.}, $\phi^{\lambda_{_N}}_{\lambda_q \lambda_d}$, are boost-invariant and rotational invariant (manifestly in the transverse plane), and thus independent of the nucleon momentum $p_{_N}$.   

\end{widetext}

The wave functions involving scalar-diquark are  
\begin{eqnarray}
\phi^{\lambda_{_N}}_{\lambda_q} &=&  \bar{u}(k,\lambda_q) \left(\varphi_1^s + \frac{\mn \gamma^+}{p_{_N}^+} \varphi_2^s \right) u(p_{_N}, \lambda_{_N})\, , \label{eqn:scalardiquark1}
\end{eqnarray}
which is  the same as in Ref.~\cite{Cloet:2012cy}, while the axial-diquark is different,
\begin{align}
	\phi^{\lambda_{_N}}_{\lambda_q \lambda_a} =  \bar{u}(k,\lambda_q) \bar{\varepsilon}^\ast_\mu(q,\lambda_a) \Big(&\varphi_1^a \gamma^\mu\gamma_5 \label{eqn:axialdiquark1} \\
	&+  \varphi_2^a \frac{q^\mu}{\mn} \gamma_5 \Big) u(p_{_N}, \lambda_{_N})\, . \notag
\end{align}
For the axial-diquark, we use the modified vector introduced in \cite{Yan:1973qg} for its $\bar{\varepsilon}_\mu$. It is related to the usual definition of a polarization vector  $\varepsilon_\mu$ (satisfying $q^\mu \varepsilon_\mu =0$), through $\bar{\varepsilon}_\mu = \varepsilon_\mu - \varepsilon^+ q_\mu/q^+ $. By choosing an appropriate frame such that $q^\mu=\left(q^+, \frac{\ma^2}{q^+}, \vec{0}_\perp \right)$, we have $\varepsilon^\mu(\lambda_a=\pm 1)=\left(0, 0, \vec{\varepsilon}(\pm 1)\right)$,  $\varepsilon^\mu(\lambda_a=0)=\left(\frac{q^+}{\ma}, -\frac{\ma}{q^+}, \vec{0}_\perp\right)$, and thus  $\bar{\varepsilon}^\mu(\lambda_a=\pm 1)=\left(0, 0, \vec{\varepsilon}(\pm 1)\right)$ and $\bar{\varepsilon}^\mu(\lambda_a=0)=\left(0, -2 \frac{\ma}{q^+}, \kperp{0}\right)$. 

The intrinsic wave functions, $\varphi_{1,2}^{s,a}$, in Eqs.~(\ref{eqn:scalardiquark1}) and~(\ref{eqn:axialdiquark1}), are scalar functions of intrinsic variables, $x$, $\kperp{k}$ (their details are discussed in Sec.~\ref{sec:input}), while $\phi^{\lambda_{_N}}_{\lambda_q \lambda_a}$ are functions of  $x$, $\kperp{k}$ and helicities of participating DOFs including diquarks and nucleon. The relationships between $\varphi$ and $\phi$ wave functions are collected in Appendix~\ref{app:quarkwf} and Tables~\ref{tab:qswf} and~\ref{tab:qawf}. 

With the wave functions set up, we first define the current matrix elements between spin states, not worrying about flavor space for the moment,  {\it e.g.}, $ ^{s.s}\langle \lambda_{_N}; N  \vert J_\mathrm{EM}^+\vert\lambda_{_N}; N \rangle^{s.s} $, and define $f_{1s}$, $f_{2s}$ as 
\begin{eqnarray}
f_{1s} &\equiv &  ^{s.s}\langle \half; N  \vert J_\mathrm{EM}^+ \vert \half; N \rangle^{s.s}  \ , \label{eqn:f1sdef}  \\ 
 -\frac{q^R}{\sqrt{2}\mn} f_{2s}  & \equiv & ^{s.s}\langle -\half; N  \vert  J_\mathrm{EM}^+ \vert \half; N \rangle^{s.s} \ , \label{eqn:f2sdef}  \\ 
 \tilde{f}_{s} & \equiv &  ^{s.s}\langle \half; N  \vert J_A^+ \vert \half; N \rangle^{s.s}   \ . \label{eqn:f1stildedef}
\end{eqnarray}
and similarly for $f_{1a}$, $f_{2a}$, and $f_{Aa}$  in terms of $\vert \lambda_{_N}; N \rangle^{s.t}$. In the above three equations, $J_\mathrm{EM}^\mu = \bar{q} \gamma^\mu q$,  $ J_A^\mu = \bar{q} \gamma^\mu \gamma_5  q $, with $q$ fixed as the struck quark. The isospin dependence will be discussed later. The $1/(2p_{_N}^+)$ is already canceled out by the overlap of CM motion state, as compared to Eqs.~(\ref{eqn:F1N})--(\ref{eqn:GAN}). By using the Lepage-Brodsky convention for the Dirac spinors~\cite{Brodsky:1997de}, we can express these quantities in terms of overlap of light-front wave functions:  
\begin{eqnarray}
f_{1s} &=&  \int\!\! d\mu \sum_{\lambda_q} \phi^{\half \ast }_{\lambda_q  }(x, \kperp{k}') \phi^{\half }_{\lambda_q  }(x, \kperp{k} ) \label{eqn:f1s} \\ 
f_{1a} &=&  \int\!\! d\mu   \sum_{\lambda_q,\lambda_a}\phi^{\half \ast }_{\lambda_q \lambda_a}(x, \kperp{k}') \phi^{\half}_{\lambda_q \lambda_a}(x, \kperp{k} ) \label{eqn:f1a}   \\
f_{2s} &=& -\frac{\sqrt{2}\mn}{q^R} \int\!\! d\mu  \sum_{\lambda_q} \phi^{-\half \ast }_{\lambda_q  }(x, \kperp{k}') \phi^{\half }_{\lambda_q  }(x, \kperp{k} )  \label{eqn:f2s} \\ 
f_{2a} &=& -\frac{\sqrt{2}\mn}{q^R}  \label{eqn:f2a} \\
&\times& \int\!\! d\mu \sum_{\lambda_q,\lambda_a}\phi^{-\half \ast }_{\lambda_q \lambda_a}(x, \kperp{k}') \phi^{\half}_{\lambda_q \lambda_a}(x, \kperp{k} )  \ , \notag 
\end{eqnarray}
for the EM current, while for the axial current we get,
\begin{eqnarray}
\tilde{f}_{s} &=&  \int\!\! d\mu \sum_{\lambda_q} (-)^{\lambda_q-\half} \phi^{\half \ast }_{\lambda_q  }(x, \kperp{k}') \phi^{\half }_{\lambda_q  }(x, \kperp{k} )  \label{eqn:fAs}  \\ 
\tilde{f}_{a} &=&  \int\!\! d\mu\!\!  \sum_{\lambda_q,\lambda_a} (-)^{\lambda_q-\half} \phi^{\half \ast }_{\lambda_q \lambda_a}(x, \kperp{k}') \phi^{\half}_{\lambda_q \lambda_a}(x, \kperp{k} ) . \ \ \ \ \label{eqn:fAa}  
\end{eqnarray}
Inside these integrands, $ d\mu \equiv \frac{d x d \vec{k}_\perp}{16\pi^3 x(1-x)}$, $\kperp{k}'=\kperp{k}+ (1-x)\kperp{q}$, $\lambda_q = \pm 1/2$ and $\lambda_a= 0, \pm 1$. It should be pointed out that
the 2nd-class axial current is zero here~\cite{Weinberg:1996kr}, because isospin symmetry is respected in this model. The detailed expression of these form factors in terms of $\varphi_{1,2}^{s,a}$ can
be found in Appendix~\ref{app:quarkwf}. 

To compute the current matrix elements with wave functions $\vert \lambda_{_P}; P \rangle_\mathrm{q\otimes d}$, we need to sum up the contributions from the struck quarks ($3$ for nucleon)  and take into account the flavor structure of the quark-diquark wave function and the charges of the struck quarks. We then get the form factor from the nucleon's bare quark-diquark core,  
\begin{eqnarray}
F_{1p}^0 &=& \frac{3}{2} e_u f_{1s} +\left(\half e_u +e_d \right) f_{1a} =f_{1s}  \label{eqn:barequarkff1} \\ 
F_{2p}^0 &=& \frac{3}{2} e_u f_{2s} +\left(\half e_u +e_d \right) f_{2a} =f_{2s}  \label{eqn:barequarkff2} \\ 
F_{1n}^0 &=& \frac{3}{2} e_d f_{1s} +\left(\half e_d +e_u \right) f_{1a} = \half f_{1a}-\half f_{1s} \label{eqn:barequarkff3} \\ 
F_{2n}^0 &=& \frac{3}{2} e_d f_{2s} +\left(\half e_d +e_u \right) f_{2a} = \half f_{2a}-\half f_{2s} \label{eqn:barequarkff4} \\ 
\tilde{F}_{1p}^0 &=& \frac{3}{2} e_{Au} \tilde{f}_{s} +\left(\half e_{Au} +e_{Ad} \right) \tilde{f}_{a} = \frac{3}{2} \tilde{f}_{s} -\half \tilde{f}_{a} \label{eqn:barequarkff5} \\
\tilde{F}_{1n}^0 &=& \frac{3}{2} e_{Ad} \tilde{f}_{s} +\left(\half e_{Ad} +e_{Au} \right) \tilde{f}_{a} = -\tilde{F}_{1p}^0\, . \label{eqn:barequarkff6} 
\end{eqnarray}
In the above expressions, $e_q$ and $e_{Aq}$ are the EM and axial charges of the quarks with the latter $e_{Aq}\! =\! \pm 1$ for the $u$- and $d$-quark, respectively. However, since the axial current is not conserved, the axial charge of a constituent quark, as employed in our model, is not expected to be exactly $\pm 1$.  Therefore, the size of $e_{Aq}$  will be adjusted later so that our predicted nucleon axial charge, $\FAn{1}(0) $, agrees with the experimental value.

\subsection{\label{subsec:pion} Pion cloud Diagram (II) and (III)}

\subsubsection{Preparations } \label{subsec:pioncloudprep}

To simplify the following presentations, a series of definitions of the EW current matrix elements and strong interaction matrix elements, {\it i.e.}, the vertices of Diagrams \II~and \III~in
Fig.~\ref{fig:FeynmannD}, need to be constructed.  The calculations of those diagrams are based on the strong interaction terms quantized on the light front: 
\begin{eqnarray} 
\label{eq:lagrange}
V_{int}= - \int d x_+ d\kperp{x}&& \bigg[ \frac{g_A}{f_\pi} \bar{N}   \gamma^\mu  \gamma_5  \partial_\mu\vec{\pi}  \frac{\vec{\tau}}{2}  N \\ 
       + &&   \frac{h_A}{f_\pi} \bar{\Delta}^a_{ \mu} T^{1,i;\half,\sigma}_{\frac{3}{2},a} \partial^\mu \pi_i N_\sigma +  \mathrm{h.c.} \bigg] \notag \\ 
   = - \int d x_+ d\kperp{x}&& \bigg[ g_{_{\pi NN}} \bar{N}    i\,\gamma_5  \vec{\pi}\cdot\vec{\tau}   N  \notag \\ 
       + &&   \frac{h_A}{f_\pi} \bar{\Delta}^a_{ \mu} T^{1,i;\half,\sigma}_{\frac{3}{2},a} \partial^\mu \pi_i N_\sigma +  \mathrm{h.c.} \bigg]
       \, . \notag
\end{eqnarray}  
Here $N$,$\Delta$, $\vec{\pi}$ are the fields of the nucleon, $\Delta$ resonance, and pion; the pion decay constant is $f_\pi \approx 94$ MeV, nucleon's axial charge $g_A =1.27 $; in the $N-\Delta-\pi$ coupling,  $a$, $\sigma$, $i$ are the isospin indices for the representations of 
isospin $3/2$, $1/2$, and $1$ multiplets; $T^{1,i; \half, A}_{\frac{3}{2},a}$ is the C-G coefficients combining isovector current and isospin $\half$ to form isospin $3/2$ \cite{Serot:2012rd}. The
pseudo-vector $N-N-\pi$ coupling is connected to the pseudo-scalar coupling for on-shell
nucleons, $ g_{_{\pi NN}} \bar{N}   i  \gamma_5  \vec{\pi}  \vec{\tau}  N $, and
$g_{_{\pi NN}} = \frac{\mn}{f_\pi} g_A \approx 13.5 $ \cite{Cloet:2012cy}. 

We use the second expression appearing in Eq.~(\ref{eq:lagrange}) because $g_{_{\pi NN}}$ more accurately represents the empirical
pion-nucleon coupling constant for on-mass-shell nucleons relative to $g_A / f_\pi$. We also emphasize that the direct $\gamma N \pi$ contact
interactions are implicitly included using the pseudoscalar Lagrangian. Moreover, among the pion-cloud contributions, the most important
piece comes from the $\gamma\pi$ interaction in Diagram ({\bf III}) of Fig.~\ref{fig:FeynmannD}. Given that this latter graph is correctly evaluated
by keeping the pole term in which the spectator nucleon is on its mass-shell, there is no important difference between the two
forms in Eq.~(\ref{eq:lagrange}) beyond the noted choice of coupling constant.

As noted above, a possible term involving a direct $aN\pi$ coupling is not included. It is possible that including the neglected term along with the
effects of using a pseudovector coupling could bring the computed value of $g_A$ (to be discussed below) into better agreement with experiment.

The matrix elements of $V_{int}$ as needed in the diagram calculations can be presented with isospin structure explicitly factorized out:
\begin{eqnarray}
\langle \lambda_{_{Nf}} ;  N^{\sigma_f}, \pi^i \vert V_{int} \vert \lambda_{_{Ni}}; N^{\sigma_i} \rangle &\equiv &  g_{_{\pi NN}} \left( \tau_i \right)_{\sigma_f}^{\ \sigma_i}   \mathcal{V}_{\lambda_{_{Nf}},\lambda_{_{Ni}}} (x,\kperp{k}) \notag \\ \\    
 \langle \lambda_{_{\Delta  }} ;  \Delta^{a }, \pi^i \vert V_{int} \vert \lambda_{_{N }}; N^{\sigma } \rangle &\equiv & \frac{h_A}{f_\pi} \delta_{ij} T^{1,j;\half,\sigma}_{\frac{3}{2},a}   \mathcal{V}_{\lambda_{_{\Delta }},\lambda_{_{N }}} (x,\kperp{k}) \notag  \\
\end{eqnarray}
Note the two sets of matrix elements are Lorentz-boost and transverse-rotation invariant; they are functions of the intrinsic kinetic variables, $x$ and $\kperp{k}$.
 These matrix elements, when multiplied by the appropriate energy denominators (see, {\it e.g.}, Ref.~\cite{Cloet:2012cy}), represent Fock-space components of the
nucleon wave function.
We compute these matrix elements, assuming the baryon in the final state in the CM frame carry momentum fraction $x$ and transverse momentum $\kperp{k}$, while the accompanying $\pi$ carrying $1-x$ and $-\kperp{k}$. This is in parallel to the assignment in the quark-diquark wave function definitions [cf.~Eqs.~(\ref{eqn:qdqdef1}) and~(\ref{eqn:qdqdef2}].  The detailed results are gathered in Table.~\ref{tab:NNpi} and~\ref{tab:NDpi} in Appendix~\ref{app:hadronicvertices}, where a few details for the calculation can also be found including the convention for spin $\frac{3}{2}$ spinor. The results are consistent with those in Ref.~\cite{Pasquini:2007iz}.

Moreover, we need to set up the convention for the current matrix elements involving $\Delta$. We use the Lorentz-covariant basis from Ref.~\cite{Pascalutsa:2006up} for the EM current and the basis from Ref.~\cite{Leitner:2008ue} for the axial current \footnote{a pure imaginary factor is absorbed into definition of $\Gamma^{\alpha \mu}_{_{\gamma N; \Delta}}$ as compared to its definition in Ref.~\cite{Pascalutsa:2006up}; and a real factor $\sqrt{3/2}$ is absorbed in $\Gamma^{\alpha \mu}_{_{A N; \Delta}}$.  See discussions in Sec.~\ref{subsec:inputpioncloud}.}: 
\begin{widetext}
\begin{eqnarray}
\langle p_{_\Delta}; \Delta^a \vert J^\mu_\mathrm{EM} \vert p_{_N}; N^\sigma \rangle & \equiv &  T^{1,i=0; \half, \sigma}_{\frac{3}{2},a}  \bar{u}_\alpha(p_{_\Delta},\lambda_{_\Delta}) \Gamma^{\alpha \mu}_{_{\gamma N; \Delta}}( q,p_{_N}; p_{_\Delta})  u(p_{_N}, \lambda_{_N})  \\ 
\langle p_{_\Delta}; \Delta^a \vert J^{i, \mu}_A \vert p_{_N}; N^\sigma \rangle & \equiv &    T^{1,i; \half, \sigma}_{\frac{3}{2},a}    \bar{u}_\alpha(p_{_\Delta},\lambda_{_\Delta}) \Gamma^{\alpha \mu}_{_{A N; \Delta}}( q,p_{_N}; p_{_\Delta})  u(p_{_N}, \lambda_{_N})  \label{eqn:NDaxialdef} \\ 
\Gamma^{\alpha \mu}_{_{\gamma N; \Delta}}( q,p_{_N}; p_{_\Delta})  &\equiv &
 i \gmd \, \varepsilon^{\alpha\mu\rho\sigma} p_{_\Delta\rho}\, q_\sigma - \ged \left(q^\alpha p_{_\Delta}^\mu - p_{_\Delta}\cdot q g^{\alpha \mu}  \right) \gamma_5 -  \gcd \left(q^\alpha q^\mu - q^2 g^{\alpha \mu}  \right) \gamma_5   \\ 
 \Gamma^{\alpha \mu}_{_{A N; \Delta}}( q,p_{_N}; p_{_\Delta})  &\equiv &
 \frac{ \ca{3}}{\mn}\left(g^{\alpha \mu} \slashed{q} -q^\alpha \gamma^\mu\right)   +  \frac{\ca{4}}{\mn^2} \left( p_{_\Delta}\cdot q g^{\alpha \mu} -q^\alpha p_{_\Delta}^\mu  \right) +  \ca{5}  g^{\alpha \mu}  +\frac{\ca{6}}{\mn^2}  q^\alpha q^\mu  
\end{eqnarray}
Here  $q\equiv p_{_\Delta}-p_{_N}$. In the expressions above, we point out the EM current's isospin projection of $i\!=\!0$; in contrast, the axial current's
isospin projection can assume values of $i\!=\!\pm,\,0$, but, in the following calculations of the axial current's matrix elements, we always take $i\!=\!0$ without
loss of generality.
That being said, in the eventual charge-current calculations shown later in this analysis, it is the $i\!=\!\pm$ spherical combinations of the axial isospin components
which are relevant.
Our convention for the Levi-Civita tensor is $\varepsilon_{0123}=1$ \cite{Pascalutsa:2006up, Brodsky:1997de}, while the metric
$g^{\mu\nu}=Diag(1,-1,-1,-1)$ \cite{Pascalutsa:2006up, Brodsky:1997de}. With this convention, under light-front quantization, $\varepsilon_{+12-}=-\half$
\footnote{This is different from the one mentioned Ref.~\cite{Brodsky:1997de}}. Then, hermiticity of $J_\mathrm{EM}^\mu$ dictates that
$\langle  p_{_N}; N^\sigma\vert J^\mu(q) \vert  p_{_\Delta}; \Delta^a \rangle = \left(\langle p_{_\Delta}; \Delta^a \vert J^\mu(-q)\vert p_{_N}; N^\sigma \rangle\right)^\ast$,
but with  $q\equiv p_{_N}-p_{_\Delta}$.

For form factors at space-like momentum transfer, {\it i.e.}, $Q^2 \geq 0$, we can always boost the system to a frame with $q^+=0$, where the matrix element of
$J^+/2p_{_N}^+$ (a Lorentz invariant) can be computed more easily. In the following, the matrix elements will be defined with the isospin structure manifestly
factorized out:  
\begin{eqnarray}
\mathcal{J}^{(0)V}_{\lambda_{_\Delta}, \lambda_{_N}} (q) &\equiv & \bar{u}_\alpha(p_{_\Delta},\lambda_{_\Delta}) \Gamma^{\alpha \mu=+}_{_{\gamma N; \Delta}}( q,p_{_N}; p_{_\Delta})  u(p_{_N}, \lambda_{_N})/(2 p_{_N}^+) \\ 
\mathcal{J}^{(0)A}_{\lambda_{_\Delta}, \lambda_{_N}} (q) &\equiv & \bar{u}_\alpha(p_{_\Delta},\lambda_{_\Delta}) \Gamma^{\alpha +}_{_{A N; \Delta}}( q,p_{_N}; p_{_\Delta})  u(p_{_N}, \lambda_{_N})/(2 p_{_N}^+)
\end{eqnarray}
Note the superscript ``V'' for the EM current is due to the fact that only the isovector component of the EM current participate in the $N\leftrightarrow \Delta$ transitions. Both quantities are functions of momentum transfer $q$. Carrying $\lambda_{_\Delta}, \lambda_{_N}$ indices suffice to indicate they are for the inelastic transition current. The results for both $\mathcal{J}^{(0)V}_{\lambda_{_\Delta}, \lambda_{_N}} (q)$ and $\mathcal{J}^{(0)A}_{\lambda_{_\Delta}, \lambda_{_N}} (q)$ are collected in Table~\ref{tab:NDEMcurrentME} and ~\ref{tab:NDAxialcurrentME}.

For the EW elastic current matrix elements of the $\Delta$-baryon, we follow the conventions in Refs.~\cite{Pascalutsa:2006up} and~\cite{Alexandrou:2013opa}:
\begin{eqnarray}
\langle p_{_\Delta}'; \Delta^{a'} \vert J^\mu_\mathrm{EM} \vert p_{_\Delta}; \Delta^a \rangle & \equiv &  \left(\half+ t^0\right)_{a'}^{\, a} \bar{u}_\alpha\left( p_{_\Delta}',  \lambda_{_\Delta}'\right) \Gamma^{\alpha \mu \beta}_{_\Delta; \gamma _\Delta}(q, p_{_\Delta}; p_{_\Delta}')  u_\beta\left( p_{_\Delta},  \lambda_{_\Delta}\right)  \notag \\ 
\langle p_{_\Delta}'; \Delta^{a'} \vert J^{0,\mu}_A \vert p_{_\Delta}; \Delta^a \rangle & \equiv & \left( t^0\right)_{a'}^{\, a}\bar{u}_\alpha\left( p_{_\Delta}',  \lambda_{_\Delta}'\right)  \Gamma^{\alpha \mu \beta}_{_\Delta; A _\Delta}(q, p_{_\Delta}; p_{_\Delta}')  u_\beta\left( p_{_\Delta},  \lambda_{_\Delta}\right)  \notag \\ 
 \Gamma^{\alpha \mu \beta}_{_\Delta; \gamma _\Delta}(q, p_{_\Delta}; p_{_\Delta}') & \equiv  & 
-\left(\Fd{1} g^{\alpha\beta} + \Fd{3}\frac{q^\alpha q^\beta}{4\md^2} \right) \gamma^\mu  -\left(\Fd{2} g^{\alpha\beta} + \Fd{4}\frac{q^\alpha q^\beta}{4\md^2} \right) \frac{\sigma^{\mu\nu}iq_\nu}{2\md} \\ 
 \Gamma^{\alpha \mu \beta}_{_\Delta; A _\Delta}(q, p_{_\Delta}; p_{_\Delta}')  & \equiv & 
-\left(\FAd{1} g^{\alpha\beta} + \FAd{3}\frac{q^\alpha q^\beta}{4\md^2} \right) \gamma^\mu\gamma_5  -\left(\FAd{2} g^{\alpha\beta} + \FAd{4}\frac{q^\alpha q^\beta}{4\md^2} \right) \frac{q^\mu}{2\md}\gamma_5
\end{eqnarray}
Here $t^0$ is the isospin group generator along the 3rd direction in the isospin $= 3/2$ representation. Again $a$ and $b$ are the isospin projection of the $\Delta$ states. For the axial current, only the first  two terms, $\FAd{1}$ and $\FAd{3}$, contribute in Diagrams \II~and \III. We separate the isospin structure and define 
\begin{eqnarray}
\mathcal{J}^{(0)\mathrm{EM}}_{\lambda_{_\Delta}', \lambda_{_\Delta}} (q) &\equiv & \bar{u}_\alpha(p_{_\Delta}',\lambda_{_\Delta}') \Gamma^{\alpha + \beta}_{_\Delta; \gamma _\Delta}(q, p_{_\Delta}; p_{_\Delta}')  u_\beta\left( p_{_\Delta},  \lambda_{_\Delta}\right)/(2 p_{_\Delta}^+) \ , \\ 
\mathcal{J}^{(0)A}_{\lambda_{_\Delta}', \lambda_{_\Delta}} (q) &\equiv & \bar{u}_\alpha(p_{_\Delta}',\lambda_{_\Delta}') \Gamma^{\alpha + \beta}_{_\Delta; {_A} {_\Delta}}(q, p_{_\Delta}; p_{_\Delta}')  u_\beta\left( p_{_\Delta},  \lambda_{_\Delta}\right)/(2 p_{_\Delta}^+) \ . \\
\end{eqnarray}
The corresponding matrix elements can be found in Table~\ref{tab:DDEMmatrix} and~\ref{tab:DDAmatrix}.

\subsubsection{Previous calculations} 
By computing Diagrams~\II~and~\III~on the light front with pion-baryon intermediate states~\cite{Brodsky:1997de, Matevosyan:2005bp, Pasquini:2007iz, Cloet:2012cy}, we get their contributions
to the nucleon form factors. The EM expressions have been derived in Ref.~\cite{Cloet:2012cy,Pasquini:2007iz}, while the axial current was also studied in Ref.~\cite{Pasquini:2007iz}. Our results
are consistent with those in Ref.~\cite{Cloet:2012cy}. Here we present them together for a self-contained discussion and pay attention to the isospin structures.  

Diagram \II~gives 
\begin{eqnarray}
F_{1}^{(IIN)} &=& \left[\frac{3}{2}\left(F_{1p}^0 + F_{1n}^0\right)\delta_{\sigma_f}^{\sigma_i} - \frac{1}{2}\left(F_{1p}^0 - F_{1n}^0\right)(\tau^0)_{\sigma_f}^{\sigma_i} \right] \mathcal{F}_{11}^{(IIN)} \\ 
	      &+& \left[\frac{3}{2}\left(F_{2p}^0 + F_{2n}^0\right)\delta_{\sigma_f}^{\sigma_i} - \frac{1}{2}\left(F_{2p}^0 - F_{2n}^0\right)(\tau^0)_{\sigma_f}^{\sigma_i} \right] \mathcal{F}_{12}^{(IIN)} \notag  \\
F_{2}^{(IIN)} &=& \left[\frac{3}{2}\left(F_{1p}^0 + F_{1n}^0\right)\delta_{\sigma_f}^{\sigma_i} - \frac{1}{2}\left(F_{1p}^0 - F_{1n}^0\right)(\tau^0)_{\sigma_f}^{\sigma_i} \right] \mathcal{F}_{21}^{(IIN)} \\
	      &+& \left[\frac{3}{2}\left(F_{2p}^0 + F_{2n}^0\right)\delta_{\sigma_f}^{\sigma_i} - \frac{1}{2}\left(F_{2p}^0 - F_{2n}^0\right)(\tau^0)_{\sigma_f}^{\sigma_i} \right] \mathcal{F}_{22}^{(IIN)} \notag    
\end{eqnarray}
with $\sigma_i$ and $\sigma_f$ as the isospin projection of the initial state and final state nucleon in current matrix element calculations, and 
\begin{eqnarray}
\mathcal{F}_{11}^{(IIN)} &=& g_{_{\pi NN}}^2 \int \frac{dx d\kperp{k}}{16\pi^3 x^2(1-x) } \frac{ \left[\kperp{k}^2 - \frac{(1-x)^2}{4}Q^2 +(1-x)^2 \mn^2 \right] F_{_{\pi NN}}\left(x, \vec{k}_{f\perp}\right) F_{_{\pi NN}}\left(x, \vec{k}_{i\perp}\right)} {\left[M_{_{\pi N}}^2(x, \vec{k}_{f\perp})-\mn^2\right]  \left[M_{_{\pi N}}^2(x, \vec{k}_{i\perp})-\mn^2\right] }  \\ 
\mathcal{F}_{12}^{(IIN)} &=& -g_{_{\pi NN}}^2 \int \frac{dx d\kperp{k}}{32\pi^3 x^2  } \frac{(1-x)Q^2\, F_{_{\pi NN}}\left(x, \vec{k}_{f\perp}\right) F_{_{\pi NN}}\left(x, \vec{k}_{i\perp}\right)  }{\left[M_{_{\pi N}}^2(x, \vec{k}_{f\perp})-\mn^2\right]  \left[M_{_{\pi N}}^2(x, \vec{k}_{i\perp})-\mn^2\right] }  \\ 
\mathcal{F}_{21}^{(IIN)} &=& - g_{_{\pi NN}}^2 \int \frac{dx d\kperp{k}}{ 8 \pi^3 x^2  } \frac{ (1-x)\mn^2\,  F_{_{\pi NN}}\left(x, \vec{k}_{f\perp}\right) F_{_{\pi NN}}\left(x, \vec{k}_{i\perp}\right)} {\left[M_{_{\pi N}}^2(x, \vec{k}_{f\perp})-\mn^2\right]  \left[M_{_{\pi N}}^2(x, \vec{k}_{i\perp})-\mn^2\right] }  \\ 
\mathcal{F}_{22}^{(IIN)} &=&  g_{_{\pi NN}}^2 \int \frac{dx d\kperp{k}}{16\pi^3 x^2 (1-x)  } \frac{\left[\kperp{k}^2 + \frac{(1-x)^2}{4}Q^2 -(1-x)^2 \mn^2 - \frac{2 (\kperp{k}\cdot\kperp{q})^2}{Q^2} \right]  F_{_{\pi NN}}\left(x, \vec{k}_{f\perp}\right) F_{_{\pi NN}}\left(x, \vec{k}_{i\perp}\right)  }{\left[M_{_{\pi N}}^2(x, \vec{k}_{f\perp})-\mn^2\right]  \left[M_{_{\pi N}}^2(x, \vec{k}_{i\perp})-\mn^2\right] }  \ . 
\end{eqnarray}
In the above equations, the $N$-$N$-$\pi$ interaction includes a form factor to regularize the loop integration:  $F_{_{\pi NN}} \large( x, \kperp{k}  \large) \equiv \exp\left(-\frac{M_{_{\pi N}}^2(x, \kperp{k})-\left(\mn+m_\pi\right)^2}{2\Lambda_N^2} \right) $ with 
$M_{_{\pi N}}^2(x, \kperp{k}) \equiv \frac{\kperp{k}^2+\mn^2 }{x} + \frac{\kperp{k}^2 + \mpi^2}{1-x}$; $\vec{k}_{i\perp} \equiv \kperp{k}-\frac{1-x}{2}\kperp{q}$, and $\vec{k}_{f\perp} \equiv \kperp{k}+\frac{1-x}{2}\kperp{q}$ are the momentum carried by nucleon line---in the $\pi$-$N$ CM frame---in the vertices on the two  sides of the current vertex [cf.~Diagram \II]. 

Similarly for the isovector axial current,  
\begin{eqnarray}
\tilde{F}_1^{(IIN)} & = & \tilde{F}_{1p}^0(Q^2) (\tau^0)_{\sigma_f}^{\sigma_i} \tilde{\mathcal{F}}_{1}^{(IIN)}   \\ 
\tilde{\mathcal{F}}_{1}^{(IIN)} &\equiv & g_{_{\pi NN}}^2 \int \frac{dx d\kperp{k}}{16\pi^3 x^2(1-x) } \frac{ \left[\kperp{k}^2 - \frac{(1-x)^2}{4}Q^2 -(1-x)^2 \mn^2 \right] F_{_{\pi NN}}\left(x, \vec{k}_{f\perp}\right) F_{_{\pi NN}}\left(x, \vec{k}_{i\perp}\right)} {\left[M_{_{\pi N}}^2(x, \vec{k}_{f\perp})-\mn^2\right]  \left[M_{_{\pi N}}^2(x, \vec{k}_{i\perp})-\mn^2\right] }  
\end{eqnarray}
Note in the EM and axial form factors' definitions, the bare quark form factors from Eqs.~(\ref{eqn:barequarkff1})--(\ref{eqn:barequarkff6}) are used. 
The requirement of gauge invariance is such that using form factors generates
contact diagrams in addition to the ``Rainbow'' graphs---Diagram \II and \III---shown in Fig.~\ref{fig:FeynmannD}.  It has been argued~\cite{Miller:2002ig}
 that the momentum dependence
in the relevant kinematic region is relatively mild, and their effect is likely to be absorbed into the fitting parameters
developed in this analysis. That being the case, we compute with the dominant contributions from the graphs shown in Fig.~\ref{fig:FeynmannD},
and leave the more complicated calculations including these additional terms to future works.

Diagram \III~with $\pi N$ intermediate states gives 

\begin{eqnarray}
F_{1,2}^{(IIIN)} &=& F_\pi(Q^2)  (\tau^0)_{\sigma_f}^{\sigma_i} \mathcal{F}_{1,2}^{(IIIN)}     
\end{eqnarray}
in which $F_\pi(Q^2)$ represents the pion's EM form factor (see Sec.~\ref{subsec:inputpioncloud}) and 
\begin{eqnarray}
\mathcal{F}_{1}^{(IIIN)} &=& g_{_{\pi NN}}^2 \int \frac{dx d\kperp{k}}{8\pi^3 x^2(1-x) } \frac{ \left[\kperp{k}^2 - \frac{x^2}{4}Q^2 +(1-x)^2 \mn^2 \right] F_{_{\pi NN}}\left(x, \vec{k}_{f\perp}\right) F_{_{\pi NN}}\left(x, \vec{k}_{i\perp}\right)} {\left[M_{_{\pi N}}^2(x, \vec{k}_{f\perp})-\mn^2\right]  \left[M_{_{\pi N}}^2(x, \vec{k}_{i\perp})-\mn^2\right] }  \\ 
\mathcal{F}_{2}^{(IIIN)} &=& g_{_{\pi NN}}^2 \int \frac{dx d\kperp{k}}{4\pi^3 x } \frac{ \mn^2 F_{_{\pi NN}}\left(x, \vec{k}_{f\perp}\right) F_{_{\pi NN}}\left(x, \vec{k}_{i\perp}\right)} {\left[M_{_{\pi N}}^2(x, \vec{k}_{f\perp})-\mn^2\right]  \left[M_{_{\pi N}}^2(x, \vec{k}_{i\perp})-\mn^2\right] }  
\end{eqnarray}
It should be emphasized that $M_{_{\pi N}}^2(x, \kperp{k} )$ and $F_{_{\pi NN}} \large( x, \kperp{k}  \large)$ are the same as defined for the results of Diagram \II, but
$\vec{k}_{i\perp} \equiv \kperp{k}+\frac{x}{2}\kperp{q}$, and $\vec{k}_{f\perp} \equiv \kperp{k}-\frac{x}{2}\kperp{q}$ in the results for Diagram \III, because the external electroweak current
transfers its momentum to $\pi$ instead of $N$. Also note that Diagram \III~does not contribute to $\FAn{1}$ .

\subsubsection{Delta contribution}

Diagram \II~with N-current-$\Delta$ configuration gives 
\begin{eqnarray}
\langle \frac{J^+_{\mathrm{EM},A}}{2p_{_{Ni}}^+} \rangle & \equiv & \frac{4}{3} \left(\tau^0\right)_{\sigma_f}^{\,\sigma_i}  \mathcal{J}_{(IIN\Delta)}^{V,A}\left(q\right)_{\lambda_{_{Nf}}, \lambda_{_{Ni}}} \notag \\ 
\mathcal{J}_{(IIN\Delta)}^{V,A}\left(q\right)_{\lambda_{_{Nf}}, \lambda_{_{Ni}}} & \equiv &   \frac{g_{_{\pi NN}} h_A}{f_\pi} \int d\mu  \underset{\lambda_{_\Delta}, \lambda_{_N}}{\sum}  \notag \\ 
\quad\quad\quad\quad\quad\quad\quad\quad\quad && \frac{ \mathcal{V}^\dagger \left(x, \vec{k}_{f\perp}\right)_{\lambda_{_{Nf}}, \lambda_{_\Delta}} \mathcal{J}^{(0)V,A}\left(q\right)_{\lambda_{_\Delta}, \lambda_{_N}} \mathcal{V}\left(x, \vec{k}_{i\perp}\right) _{\lambda_{_N}, \lambda_{_{Ni}}} 
   F_{_{\pi N\Delta}}\left(x, \vec{k}_{f\perp}\right) F_{_{\pi NN}}\left(x, \vec{k}_{i\perp}\right)} {\left[M_{_{\pi \Delta}}^2(x, \vec{k}_{f\perp})-\mn^2\right]  \left[M_{_{\pi N}}^2(x, \vec{k}_{i\perp})-\mn^2\right] }  \ .  \label{eqn:JVAIIND}
\end{eqnarray}
Meanwhile, Diagram \II~with the $\Delta$-current-N configuration yields
\begin{eqnarray}
\langle \frac{ J^+_{\mathrm{EM},A}}{2p_{_{Ni}}^+} \rangle & = & \frac{4}{3} \left(\tau^0\right)_{\sigma_f}^{\,\sigma_i} \mathcal{J}_{(II\Delta N)}^{V,A}\left(q \right)_{\lambda_{_{Nf}}, \lambda_{_{Ni}}} \notag \\ 
\mathcal{J}_{(II\Delta N)}^{V,A}\left(q \right)_{\lambda_{_{Nf}}, \lambda_{_{Ni}}} & \equiv &
\frac{g_{_{\pi NN}}  h_A}{f_\pi} \int d\mu   \underset{\lambda_{_\Delta}, \lambda_{_N}}{\sum}  \notag \\ 
\quad\quad\quad\quad\quad\quad\quad\quad\quad && \frac{ \mathcal{V}^\dagger \left(x, \vec{k}_{f\perp}\right)_{\lambda_{_{Nf}}, \lambda_{_{N}}} \mathcal{J}^{(0)V,A\dagger}\left(-q \right)_{\lambda_{_N} \lambda_{_\Delta} } \mathcal{V}\left(x, \vec{k}_{i\perp}\right) _{\lambda_{_\Delta}, \lambda_{_{Ni}}} 
   F_{_{\pi NN}}\left(x, \vec{k}_{f\perp}\right) F_{_{\pi N\Delta}}\left(x, \vec{k}_{i\perp}\right)} {\left[M_{_{\pi N}}^2(x, \vec{k}_{f\perp})-\mn^2\right]  \left[M_{_{\pi \Delta}}^2(x, \vec{k}_{i\perp})-\mn^2\right] }   \label{eqn:JVAIIDN} 
\end{eqnarray}
In the two results above, another form factor for the $N$-$\Delta$-$\pi$ interaction has been introduced: 
$F_{_{\pi N \Delta}} \large( x, \kperp{k}  \large) \equiv \exp\left(-\frac{M_{_{\pi \Delta}}^2(x, \kperp{k})-\left(\md + m_\pi \right)^2}{2\Lambda_\Delta^2} \right) $
with  $M_{_{\pi \Delta}}^2(x, \kperp{k}) \equiv \frac{\kperp{k}^2+\md^2 }{x} + \frac{\kperp{k}^2 + \mpi^2}{1-x}$.
Moreover, $F_{_{\pi N N}}$, $\vec{k}_{i\perp} \equiv \kperp{k}-\frac{1-x}{2}\kperp{q}$, and
$\vec{k}_{f\perp} \equiv \kperp{k}+\frac{1-x}{2}\kperp{q}$, are the same as those for Diagram \II~with the $N$-current-$N$
configuration. 
Now let's define quantities with the isospin structure factorized away, 
\begin{eqnarray}
\mathcal{F}_{1}^{(IIN\Delta)} &= &   \mathcal{J}_{(IIN\Delta)}^{V}\left(q\right)_{ { \half} , {\half} } + \mathcal{J}_{(II\Delta N)}^{V}\left(q\right)_{ { \half} , {\half} }  \ ,  \\ 
\mathcal{F}_{2}^{(IIN\Delta)} & = &  (-)\frac{\sqrt{2}\mn}{q^R}\left[\mathcal{J}_{(IIN\Delta)}^{V}\left(q\right)_{ -\half ,  \half } +\mathcal{J}_{(II\Delta N)}^{V}\left(q\right)_{ -\half ,  \half } \right]  \ , \\ 
\tilde{\mathcal{F}}_{1}^{(IIN\Delta)} &= &   \mathcal{J}_{(IIN\Delta)}^{A}\left(q\right)_{ { \half} , {\half} } + \mathcal{J}_{(II\Delta N)}^{A}\left(q\right)_{ { \half} , {\half} }   \ . 
\end{eqnarray}
Then Diagram \II~with both the $N\Delta$ and $\Delta N$ configurations contributes to the nucleon form factors as 
\begin{eqnarray}
F_{1,2}^{(IIN\Delta)} & = & \frac{4}{3} \left(\tau^0\right)_{\sigma_f}^{\,\sigma_i}  \mathcal{F}_{1,2}^{(IIN\Delta)} \ ,  \\ 
\tilde{F}_{1}^{(IIN\Delta)} & = & \frac{8}{3} \left(\tau^0\right)_{\sigma_f}^{\,\sigma_i}  \tilde{\mathcal{F}}_{1}^{(IIN\Delta)}  \ . 
\end{eqnarray}

Now for Diagram \II~with the $\Delta$-current-$\Delta$ configuration, the matrix elements are 
\begin{align}
\langle \frac{J^+_{\mathrm{EM}}}{{2p_{_{Ni}}^+}} \rangle  = & \left(\delta_{\sigma_f}^{\,\sigma_i} + \frac{5}{3} \left(\tau^0\right)_{\sigma_f}^{\,\sigma_i}\right)    \mathcal{J}_{(II\Delta\Delta)}^{\mathrm{EM}}\left(q\right)_{\lambda_{_{Nf}}, \lambda_{_{Ni}}}   \notag \\ 
\mathcal{J}_{(II\Delta\Delta)}^{\mathrm{EM}}\left(q\right)_{\lambda_{_{Nf}}, \lambda_{_{Ni}}} \equiv &
\left(\frac{h_A}{f_\pi}\right)^2 \int d\mu \sum_{\lambda_{_\Delta}, \lambda_{_\Delta}'} &  \notag  \\
& \frac{ \mathcal{V}^\dagger \left(x, \vec{k}_{f\perp}\right)_{\lambda_{_{Nf}}, \lambda_{_\Delta}'} \mathcal{J}^{(0)EM}\left(q\right)_{\lambda_{_\Delta}', \lambda_{_\Delta}} \mathcal{V}\left(x, \vec{k}_{i\perp}\right) _{\lambda_{_\Delta}, \lambda_{_{Ni}}} 
   F_{_{\pi N\Delta}}\left(x, \vec{k}_{f\perp}\right) F_{_{\pi N\Delta}}\left(x, \vec{k}_{i\perp}\right)} {\left[M_{_{\pi \Delta}}^2(x, \vec{k}_{f\perp})-\mn^2\right]  \left[M_{_{\pi \Delta}}^2(x, \vec{k}_{i\perp})-\mn^2\right] }  \ , \label{eqn:DDcontributionEM}
\end{align}
for the EM current and 
\begin{align}
\langle \frac{J^+_{A}}{{2p_{_{Ni}}^+}} \rangle  = &  \frac{5}{3} \left(\tau^0\right)_{\sigma_f}^{\,\sigma_i} \mathcal{J}_{(II\Delta\Delta)}^{A}\left(q\right)_{\lambda_{_{Nf}}, \lambda_{_{Ni}}} \notag \\
\mathcal{J}_{(II\Delta\Delta)}^{A}\left(q\right)_{\lambda_{_{Nf}}, \lambda_{_{Ni}}}  \equiv &
\left(\frac{h_A}{f_\pi}\right)^2 \int d\mu \sum_{\lambda_{_\Delta}, \lambda_{_\Delta}'} &  \notag \\ 
&\frac{ \mathcal{V}^\dagger \left(x, \vec{k}_{f\perp}\right)_{\lambda_{_{Nf}}, \lambda_{_\Delta}'} \mathcal{J}^{(0)A}\left(q\right)_{\lambda_{_\Delta}', \lambda_{_\Delta}} \mathcal{V}\left(x, \vec{k}_{i\perp}\right) _{\lambda_{_\Delta}, \lambda_{_{Ni}}} 
   F_{_{\pi N\Delta}}\left(x, \vec{k}_{f\perp}\right) F_{_{\pi N\Delta}}\left(x, \vec{k}_{i\perp}\right)} {\left[M_{_{\pi \Delta}}^2(x, \vec{k}_{f\perp})-\mn^2\right]  \left[M_{_{\pi \Delta}}^2(x, \vec{k}_{i\perp})-\mn^2\right] }  \ , \label{eqn:DDcontributionA}
\end{align}
for the axial current. The definition of $F_{_{\pi N \Delta}}$, $\vec{k}_{i\perp} \equiv \kperp{k}-\frac{1-x}{2}\kperp{q}$, and $\vec{k}_{f\perp} \equiv \kperp{k}+\frac{1-x}{2}\kperp{q}$, are the same as those for Diagram \II~with the N-current-$\Delta$ configuration.  After defining 
\begin{eqnarray}
\mathcal{F}_{1}^{(II\Delta\Delta)} &= &   \mathcal{J}_{(II\Delta\Delta)}^{\mathrm{EM}}\left(q\right)_{ { \half} , {\half} } \ ,  \\ 
\mathcal{F}_{2}^{(II\Delta\Delta)} & = &  (-)\frac{\sqrt{2}\mn}{q^R} \mathcal{J}_{(II\Delta\Delta)}^{\mathrm{EM}}\left(q\right)_{ -\half ,  \half } \ ,  \\ 
\tilde{\mathcal{F}}_{1}^{(II\Delta\Delta)} &= &   \mathcal{J}_{(II\Delta\Delta)}^{A}\left(q\right)_{ { \half} , {\half} }  \ , 
\end{eqnarray}
the contribution of Diagram \II~with the $\Delta\Delta$ configuration to the form factors can be written as 
\begin{eqnarray}
F_{1,2}^{(II\Delta\Delta)} & = & \left(\delta_{\sigma_f}^{\,\sigma_i} + \frac{5}{3} \left(\tau^0\right)_{\sigma_f}^{\,\sigma_i}\right)   \mathcal{F}_{1,2}^{(II\Delta\Delta)} \ , \\ 
\tilde{F}_{1}^{(II\Delta\Delta)} & = & \frac{10}{3} \left(\tau^0\right)_{\sigma_f}^{\,\sigma_i}  \tilde{\mathcal{F}}_{1}^{(II\Delta\Delta)}   \ . 
\end{eqnarray}

For Diagram \III~with a $\Delta$-baryon in the intermediate state, 
\begin{eqnarray}
\langle \frac{J^+_{\mathrm{EM}}}{{2p_{_{Ni}}^+}} \rangle & = & -\frac{2}{3}  \left(\tau^0\right)_{\sigma_f}^{\,\sigma_i} F_\pi(Q^2)  \mathcal{J}_{(III\Delta)}^{V}\left(q\right)_{\lambda_{_{Nf}}, \lambda_{_{Ni}}}  \ ,  \\ 
\mathcal{J}_{(III\Delta)}^{V}\left(q\right)_{\lambda_{_{Nf}}, \lambda_{_{Ni}}}  & \equiv & 
\left(\frac{h_A}{f_\pi}\right)^2 \int d\mu \frac{ \sum_{\lambda_{_\Delta}}\mathcal{V}^\dagger \left(x, \vec{k}_{f\perp}\right)_{\lambda_{_{Nf}}, \lambda_{_\Delta} }  \mathcal{V}\left(x, \vec{k}_{i\perp}\right) _{\lambda_{_\Delta}, \lambda_{_{Ni}}} 
   F_{_{\pi N\Delta}}\left(x, \vec{k}_{f\perp}\right) F_{_{\pi N\Delta}}\left(x, \vec{k}_{i\perp}\right)} {\left[M_{_{\pi \Delta}}^2(x, \vec{k}_{f\perp})-\mn^2\right]  \left[M_{_{\pi \Delta}}^2(x, \vec{k}_{i\perp})-\mn^2\right] }   \ . 
\end{eqnarray}
Here, $M_{_{\pi \Delta}}^2(x, \kperp{k} )$ and $F_{_{\pi N\Delta}} \large( x, \kperp{k}  \large)$ as for the Diagram \II~results, but $\vec{k}_{i\perp} \equiv \kperp{k}+\frac{x}{2}\kperp{q}$, and $\vec{k}_{f\perp} \equiv \kperp{k}-\frac{x}{2}\kperp{q}$ are different. We can then define 
\begin{eqnarray}
\mathcal{F}_{1}^{(III\Delta)} &= &   \mathcal{J}_{(III\Delta)}^{V}\left(q\right)_{ { \half} , {\half} } \ ,  \\ 
\mathcal{F}_{2}^{(III\Delta)} & = &  (-)\frac{\sqrt{2}\mn}{q^R} \mathcal{J}_{(III\Delta)}^{V}\left(q\right)_{ -\half ,  \half } \ . 
\end{eqnarray}
Diagram \III  in the $\Delta$-current-$\Delta$ configuration contributes to the nucleon form factors as 
\begin{eqnarray}
F_{1,2}^{(III\Delta)} & = & -\frac{2}{3}  \left(\tau^0\right)_{\sigma_f}^{\,\sigma_i} F_\pi(Q^2)  \mathcal{F}_{1,2}^{(III\Delta)} 
\end{eqnarray}
Note this diagram doesn't contribute to the axial current form factor. 

After summing over all the diagrams, we have 
\begin{eqnarray}
F_{1p}&=& Z F_{1p}^0 + \left(F_{1p}^0+ 2 F_{1n}^0\right)\mathcal{F}_{11}^{(IIN)}+ \left(F_{2p}^0+ 2 F_{2n}^0\right)\mathcal{F}_{12}^{(IIN)} + F_\pi\, \mathcal{F}^{(IIIN)}_1   + \frac{4}{3} \mathcal{F}_1^{(IIN\Delta)}  + \frac{8}{3} \mathcal{F}_1^{(II\Delta\Delta)} - \frac{2}{3} F_\pi\, \mathcal{F}_1^{(III\Delta)}, \label{eqn:sumf1p} \notag \\   
F_{1n}&=& Z F_{1n}^0 + \left(F_{1n}^0+ 2 F_{1p}^0\right)\mathcal{F}_{11}^{(IIN)}+ \left(F_{2n}^0+ 2 F_{2p}^0\right)\mathcal{F}_{12}^{(IIN)} - F_\pi\, \mathcal{F}^{(IIIN)}_1    - \frac{4}{3} \mathcal{F}_1^{(IIN\Delta)} - \frac{2}{3} \mathcal{F}_1^{(II\Delta\Delta)} +  \frac{2}{3} F_\pi\, \mathcal{F}_1^{(III\Delta)},  \label{eqn:sumf1n} \notag \\ 
F_{2p}&=& Z F_{2p}^0 + \left(F_{1p}^0+ 2 F_{1n}^0\right)\mathcal{F}_{21}^{(IIN)}+ \left(F_{2p}^0+ 2 F_{2n}^0\right)\mathcal{F}_{22}^{(IIN)} + F_\pi\, \mathcal{F}^{(IIIN)}_2   + \frac{4}{3} \mathcal{F}_2^{(IIN\Delta)}  + \frac{8}{3} \mathcal{F}_2^{(II\Delta\Delta)} - \frac{2}{3} F_\pi\, \mathcal{F}_2^{(III\Delta)},   \label{eqn:sumf2p} \notag \\   
F_{2n}&=& Z F_{2n}^0 + \left(F_{1n}^0+ 2 F_{1p}^0\right)\mathcal{F}_{21}^{(IIN)}+ \left(F_{2n}^0+ 2 F_{2p}^0\right)\mathcal{F}_{22}^{(IIN)} - F_\pi\, \mathcal{F}^{(IIIN)}_2    - \frac{4}{3} \mathcal{F}_2^{(IIN\Delta)} - \frac{2}{3} \mathcal{F}_2^{(II\Delta\Delta)} +  \frac{2}{3} F_\pi\, \mathcal{F}_2^{(III\Delta)},  \label{eqn:sumf2n} \notag  \\
\tilde{F}_{1p}&=& Z \tilde{F}_{1p}^0 +  \tilde{F}_{1p}^0 \tilde{\mathcal{F}}_1^{(IIN)}  + \frac{8}{3} \tilde{\mathcal{F}}_1^{(IIN\Delta)}  + \frac{10}{3} \tilde{\mathcal{F}}_1^{(II\Delta\Delta)},  \label{eqn:sumfa1p}  \notag \\ 
\tilde{F}_{1n}&=& -\tilde{F}_{1p}\ . \label{eqn:sumFF}
\end{eqnarray}
\end{widetext}

\section{\label{sec:input} Model inputs }
This section summarizes the inputs we used for various components in our model, including for quark-diquark Fock space wave functions and for Baryon-$\pi$ Fock space wave functions. 

\subsection{The quark-diquark wave function}
We consider the quark-diquark wave functions [cf.~Eqs.~\ref{eqn:scalardiquark1} and~\ref{eqn:axialdiquark1} ] depending only on the invariant mass of the quark-diquark system, through a modified Gaussian form~\cite{Brodsky:1997de}, 
\begin{widetext}
\begin{eqnarray}
\varphi_{i}^s &=&\left[c^s_{i0}+c^s_{i1} \frac{ M_{qs}^2-(\mq + \ms)^2}{\mn^2} \right]  \exp{\left[-\frac{ M_{qs}^2-(\mq + \ms)^2 }{\beta_{si}^2} \right]} \\ 
\varphi_{i}^a &=& \left[c^a_{i0}+c^a_{i1} \frac{ M_{qa}^2-(\mq + \ma)^2}{\mn^2} \right]   \exp{\left[-\frac{ M_{qa}^2-(\mq + \ma)^2 }{\beta_{a i}^2} \right]} 
\end{eqnarray}
\end{widetext}
with $i=1,2$ and 
\begin{eqnarray}
M_{qs}^2 & \equiv & \frac{\kperp{k}^2 + \mq^2}{x}   + \frac{\kperp{k}^2 + \ms^2}{1-x} \\ 
M_{qa}^2 & \equiv & \frac{\kperp{k}^2 + \mq^2}{x}   + \frac{\kperp{k}^2 + \ma^2}{1-x} 
\end{eqnarray}
Note we can always pull out the overall normalization factor such that $c^s_{10}=c^a_{10}=1$; the normalization factors are not shown explicitly here but always
implemented in our numerical calculation. Naively, we consider the dimensionful quantities, such as $\mq$, $\ms$, and $\ma$, $\beta_{s1,2}$, and $\beta_{a1,2}$
to be at typical hadronic scale, {\it i.e.}, $\mathcal{O}(\mathrm{GeV})$, while dimensionless parameters, including $c^{s,a}_{1,1}$, $c^{s,a}_{2,0}$, and
$c^{s,a}_{2,1}$, to be $\mathcal{O}(1)$.

\subsection{Pion-cloud contributions} \label{subsec:inputpioncloud}
In the pion-cloud contributions, as shown in Eqs.~(\ref{eqn:sumf1p}), Diagram \II~with nucleon and pion intermediate states
depend on nucleon bare form factors constructed from nucleon's quark-diquark wave functions. Diagram \III~with either nucleon
or $\Delta$ intermediate states, which only contribute to the EM form
factors,  is proportional to the pion's EM form factors, $F_\pi (Q^2)$. For this quantity, we chose
$F_\pi (Q^2)\! =\! \left(1+ Q^2/\Lambda^2_\pi\right)^{-1}$, as done in Ref.~\cite{Cloet:2012cy},
which successfully used the same form, taking $\Lambda^2_\pi = 0.5~\mathrm{GeV}^2$ as also done here. 
This selection provides a robust description of the pion form factor in the kinematical region of
greatest relevance to the present study ($Q^2\! \le\! 1\,\mathrm{GeV}^2$) while similarly
agreeing with a range of experimental data at both low \cite{Amendolia:1984nz,Amendolia:1986wj} and somewhat higher
\cite{Horn:2006tm,Blok:2008jy,Huber:2008id} values of $Q^2$, all of which favor values of $\Lambda^2_\pi$ similar to our
choice of $0.5~\mathrm{GeV}^2$.

For Diagram \II~with $\Delta$(s) in the intermediate state, the same type of  quark-diquark wave functions in principle can be constructed for the $\Delta$, which dictates its bare $N\rightarrow\Delta$ inelastic and elastic form factors. However, to simplify the current work, we instead use the physical form factors  to approximately take into account their contributions. Since the $\Delta$ contribution plays a minor role in the full form factors, we expect its error to be less relevant than the error due to the uncertainty in the nucleon's quark wave function. A full and consistent study of this will be left for the future investigation. Inside  $\Delta$'s contribution, {\it e.g.}, $\mathcal{J}^{V,A}_{IIN\Delta}$ and $\mathcal{J}^{V,A}_{II\Delta N}$ (cf.~Eqs.~(\ref{eqn:JVAIIND}) and (\ref{eqn:JVAIIDN}) and Tables~\ref{tab:NDEMcurrentME} and~\ref{tab:NDAxialcurrentME}),  we need inputs for transition form factors $\ged$, $\gmd$, and $\gcd$ to compute the diagram's contribution to the nucleon EM current, and $\ca{3,4,5}$ to the axial form factor. 

For $\ged$, $\gmd$, and $\gcd$, we use information extracted from the measurements of electroproduction and photoproduction of pions ~\cite{Tiator:2011pw}: 
\begin{eqnarray}
\gmd &=& \frac{3}{2\mn Q_+} \left[\ashgmd  -   \ashged	 \right]  \\ 
\gcd &=& \frac{3}{2\mn Q_+} \left[\frac{4\md^2}{Q_-^2} \ashged + \frac{Q^2+\mn^2-\md^2}{Q_-^2}  \ashgcd \right] \notag \\  
\ged &=& \frac{3}{2\mn Q_+} \left[\frac{2 (Q^2+\mn^2-\md^2)}{Q_-^2} \ashged  -  \frac{2 Q^2}{Q_-^2} \ashgcd \right]\, , \notag
\end{eqnarray}
with $Q_\pm \equiv \sqrt{Q^2 + \left(\md \pm \mn \right)^2}$; we also use the parametrization of the Ash form factors in Ref.~\cite{Tiator:2011pw},
\begin{eqnarray}
G_{_{N\Delta}}^{\scaleto{E}{4pt}\scaleto{(M)}{6pt}}  &= & g_{_{E(M) \Delta}} \left(1 +\beta_{_{E(M)}} Q^2 \right) e^{-\gamma_{_{E(M)}} Q^2} G_D(Q^2) \\ 
\ashgcd & = & g_{_{C \Delta}} \frac{1 +\beta_{_{C}} Q^2 }{1+d_{_C} \frac{Q^2}{4 \mn^2}} \frac{4\md^2}{\md^2-\mn^2}  e^{-\gamma_{_{C}} Q^2} G_D(Q^2) . \notag \ \ \ \ 
\end{eqnarray}
The coefficients involved in the parametrizations are given in Table~\ref{tab:Tiator} \footnote{Our definition of $\gmd$, $\gcd$, $\ged $ differ from the corresponding ones in Ref.~\cite{Pascalutsa:2006up} by absorbing the factor $\frac{3(\mn+\md)}{2 \mn Q_+^2}$ into these form factors.}; $G_D(Q^2)\equiv \left[1+Q^2/(0.71\mathrm{GeV}^2)\right]^{-2}$. 

\begin{table}
 \begin{ruledtabular}
\begin{tabular}{cccc}
  & M1 & E2 &  C2 \\ \hline
$g_\alpha$ & 3 & 0.0637 & 0.124 \\ \hline 
$ \beta_\alpha (\mathrm{GeV}^{-2}) $ &  0.0095 & -0.0206 & 0.120 \\ \hline 
 $ \gamma_\alpha (\mathrm{GeV}^{-2}) $ &   0.23 & 0.16 & 0.23 \\ \hline
$ d_\alpha $ & 0 & 0 & 4.9 \\
\end{tabular} \caption{ Parameter values used for the Ash form factors  from Ref.~\cite{Tiator:2011pw}.} \label{tab:Tiator}
\end{ruledtabular}
\end{table}

For the axial-transition form factors, the Adler parametrization appearing in Ref.~\cite{Leitner:2008ue} is used, 
\begin{eqnarray}
\ca{3}(Q^2) & = & 0 \ , \notag \\ 
\ca{5}(Q^2) &=& \sqrt{\frac{3}{2}} \left[1.17\left(1- \frac{0.25 Q^2}{(0.04 +Q^2)}\right) \left(1+\frac{Q^2}{0.95^{^2}}\right)^{-2} \right]  \ , \notag \\ 
\ca{4}(Q^2) &= & -\frac{\ca{5}(Q^2)}{4} \ . 
\end{eqnarray}
The factor $\sqrt{\frac{3}{2}}$ is due to the definition of the isospin structure in Eq.~(\ref{eqn:NDaxialdef}). 

For the $\Delta$ elastic form factors needed in the calculation of Diagram \III~(see Eqs.~(\ref{eqn:DDcontributionEM}) and~(\ref{eqn:DDcontributionA}), and Tables~\ref{tab:DDEMmatrix} 
and~\ref{tab:DDAmatrix}), information is limited. The results from existing LQCD calculations~\cite{Alexandrou:2008bn} are implemented: 
\begin{eqnarray}
\Fd{1} &=& \frac{\gddezero}{\tau +1}-\frac{2 \gddetwo \tau }{3 (\tau +1)}+\frac{ \gddmone \tau }{\tau +1}-\frac{4 \gddmthree\tau
   ^2}{5 (\tau +1)} \\ 
\Fd{2} &=&   -\frac{\gddezero }{\tau +1}+\frac{2\tau \gddetwo  }{3 (\tau +1)}+\frac{\gddmone}{\tau +1}-\frac{4\tau \gddmthree
    }{5 (\tau +1)} \\ 
\Fd{3} &=& \frac{2 \gddezero}{(\tau +1)^2}-\frac{2 (2 \tau +3) \gddetwo}{3 (\tau +1)^2}+\frac{2 \tau \gddmone
   }{(\tau +1)^2}\notag \\ 
    && -\frac{2  \tau  (4 \tau +5) \gddmthree}{5 (\tau +1)^2} \\ 
 \Fd{4} &=& -\frac{2 \gddezero}{(\tau +1)^2}+\frac{2 (2
   \tau +3) \gddetwo }{3 (\tau +1)^2}+\frac{2 \gddmone}{(\tau +1)^2}\notag \\ 
   && -\frac{2 (4 \tau +5) \gddmthree}{5 (\tau +1)^2}
\end{eqnarray}
with $\tau \equiv \frac{Q^2}{4\mn^2}$. We set $\gddetwo=\gddmthree = 0 $, and 
\begin{eqnarray}  
\gddezero & = & \left(1+\frac{Q^2}{1.065^{^2}} \right)^{-2} \\ 
 \gddmone & = & 3.12 \exp{\left(-\frac{Q^2}{0.9235^{^2}}\right)} 
\end{eqnarray}       
The above parametrizations are the fits to the $m_\pi = 353 $ MeV results in Ref.~\cite{Alexandrou:2008bn}. 
For the $\Delta$'s axial elastic form factors, we  use the given parametrizations for the ``$m_\pi =0.411$ GeV with Quenched Wilson fermions'' results in Ref.~\cite{Alexandrou:2013opa} (see Table III and VI therein), 
\begin{eqnarray}
\FAd{1}& = & \frac{(0.40+ 1.98 Q^2)}{\left(Q^2+0.94^2 \right)^3} \ , \\
\FAd{3}& = & \FAd{1} \frac{3.8 }{Q^2+0.1^2 } \ . 
\end{eqnarray}       
     
Finally, all the pion-cloud diagrams involve strong-interaction form factors (cf.~the definitions in Sec.~\ref{subsec:pion}):     
\begin{eqnarray}     
F_{_{\pi NN}}\left(x, \kperp{k} \right) & = & \exp{\left(-\frac{M_{_{\pi N}}^2 -(\mn+m_\pi)^2}{2 \Lambda_N^2}\right) } \\  
F_{_{\pi N\Delta}}\left(x, \kperp{k} \right) & = & \exp{\left(-\frac{M_{_{\pi \Delta}}^2 -(\md+m_\pi)^2}{2 \Lambda_\Delta^2} \right) } 
\end{eqnarray}
with unknown $\Lambda_{N,\Delta}$. 

In short summary, we have 15 unknown parameters, including $\mq$, $\ms$, $\ma$, $c_{11}^s$, $\beta_{s1}$, $c_{20}^s$, $c_{21}^s$, $\beta_{s2}$, $c_{11}^a$, $\beta_{a1}$, $c_{20}^a$, $c_{21}^a$, $\beta_{a2}$, $\Lambda_N$, and $\Lambda_\Delta$, which need to be calibrated against experiment data.

\section{\label{sec:constraints} Model calibrations and predictions} 
\begin{figure}
\includegraphics[width=0.4 \textwidth, angle=0]{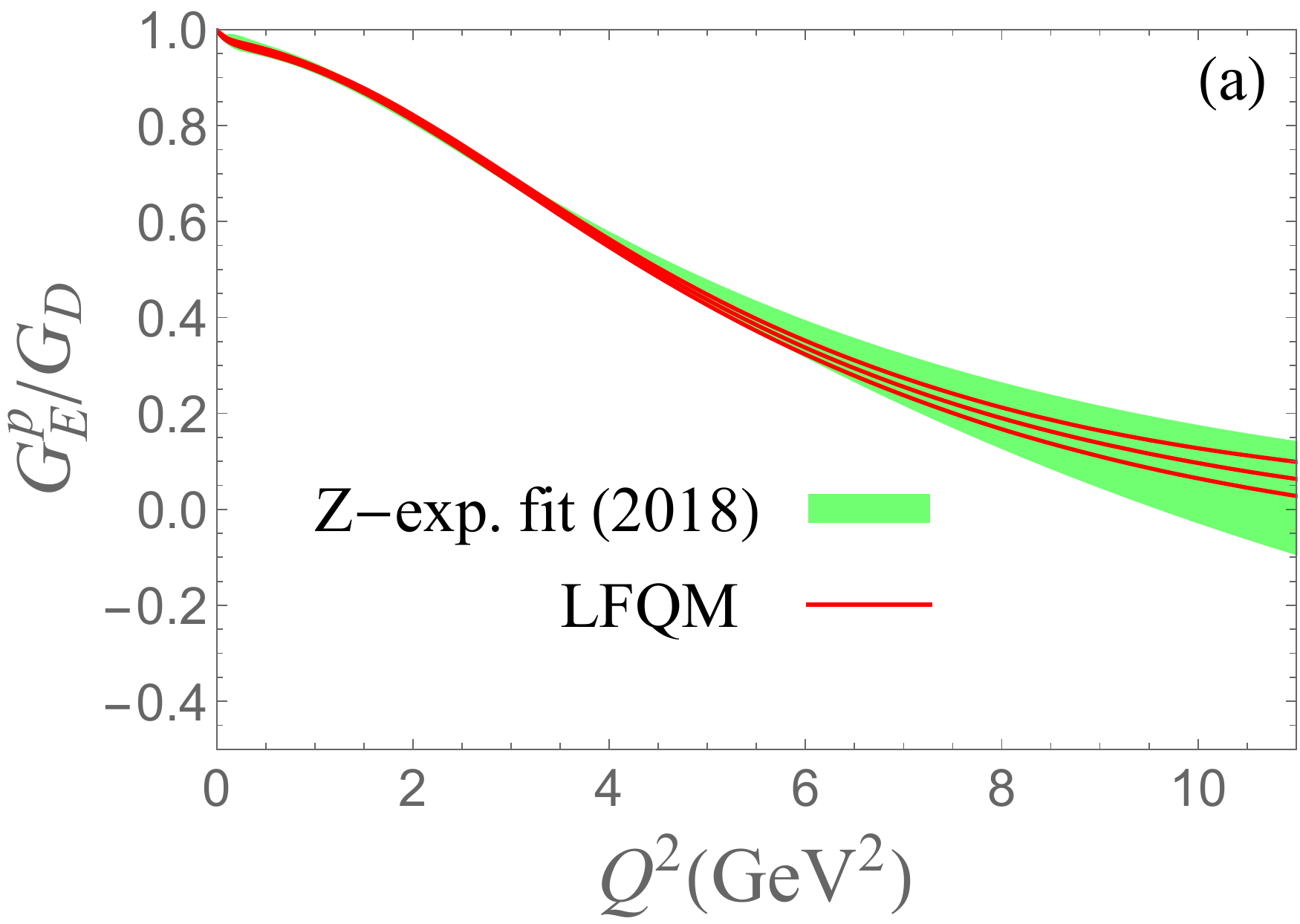} 
\includegraphics[width=0.4 \textwidth, angle=0]{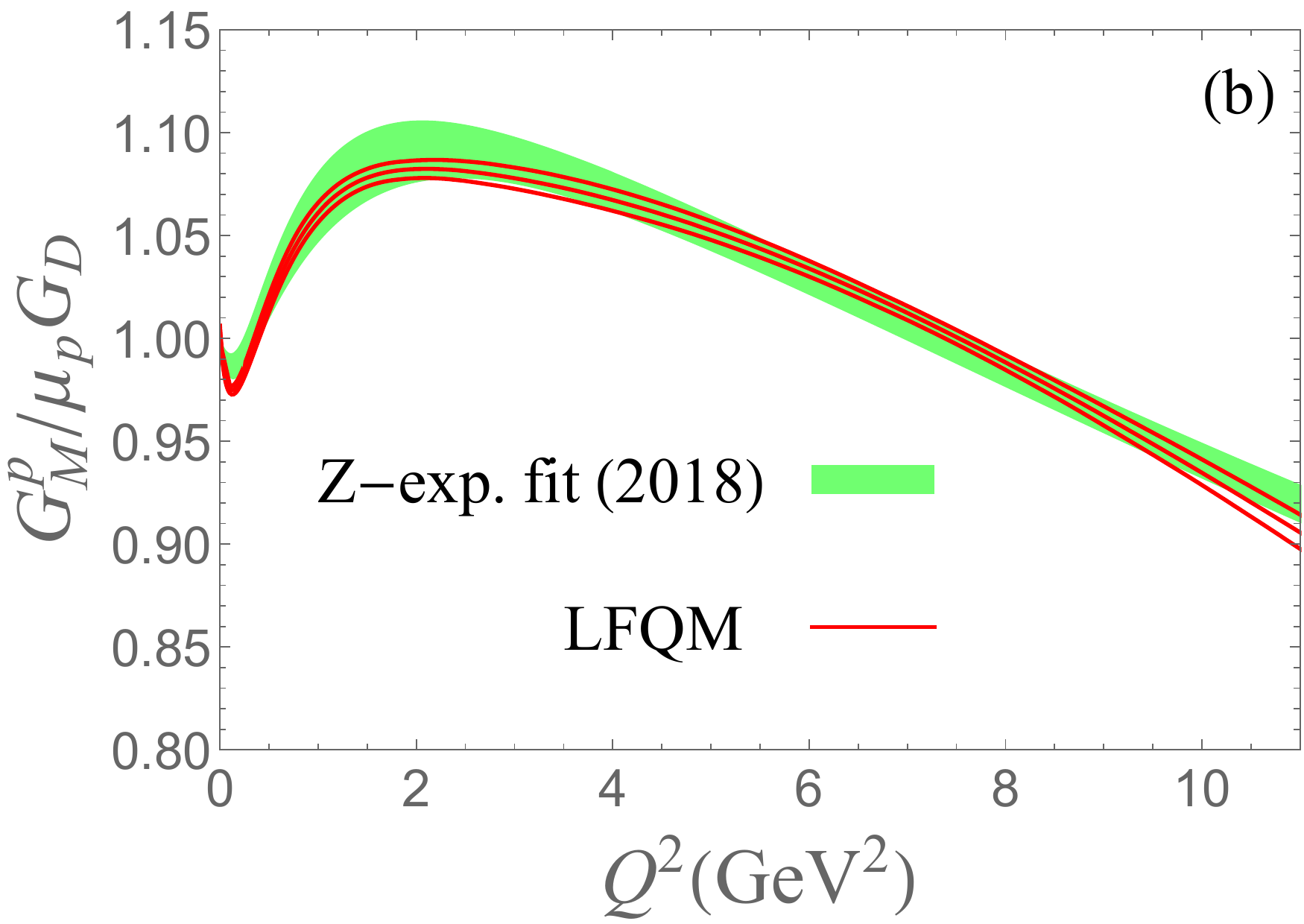}
\caption{\label{fig:Gp} Proton electric and magnetic form factors. The green band is 1-$\sigma$ error band of the results from Ref.~\cite{Ye:2017gyb} with its central value somewhere in the middle of the band. The three red solid curves are the central value and error band of our model results.}
\end{figure}

\begin{figure}
\includegraphics[width=0.4 \textwidth, angle=0]{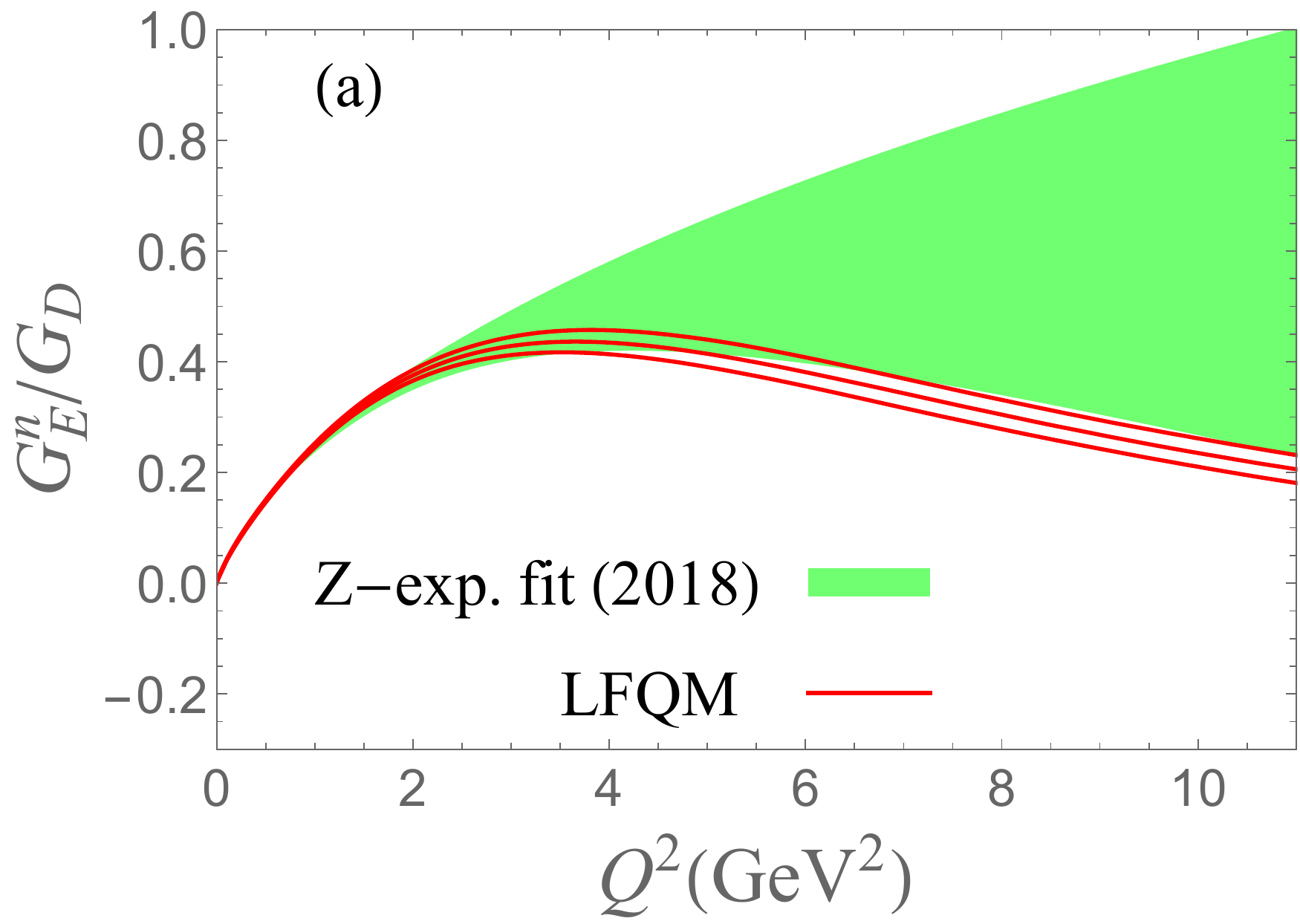} 
\includegraphics[width=0.4 \textwidth, angle=0]{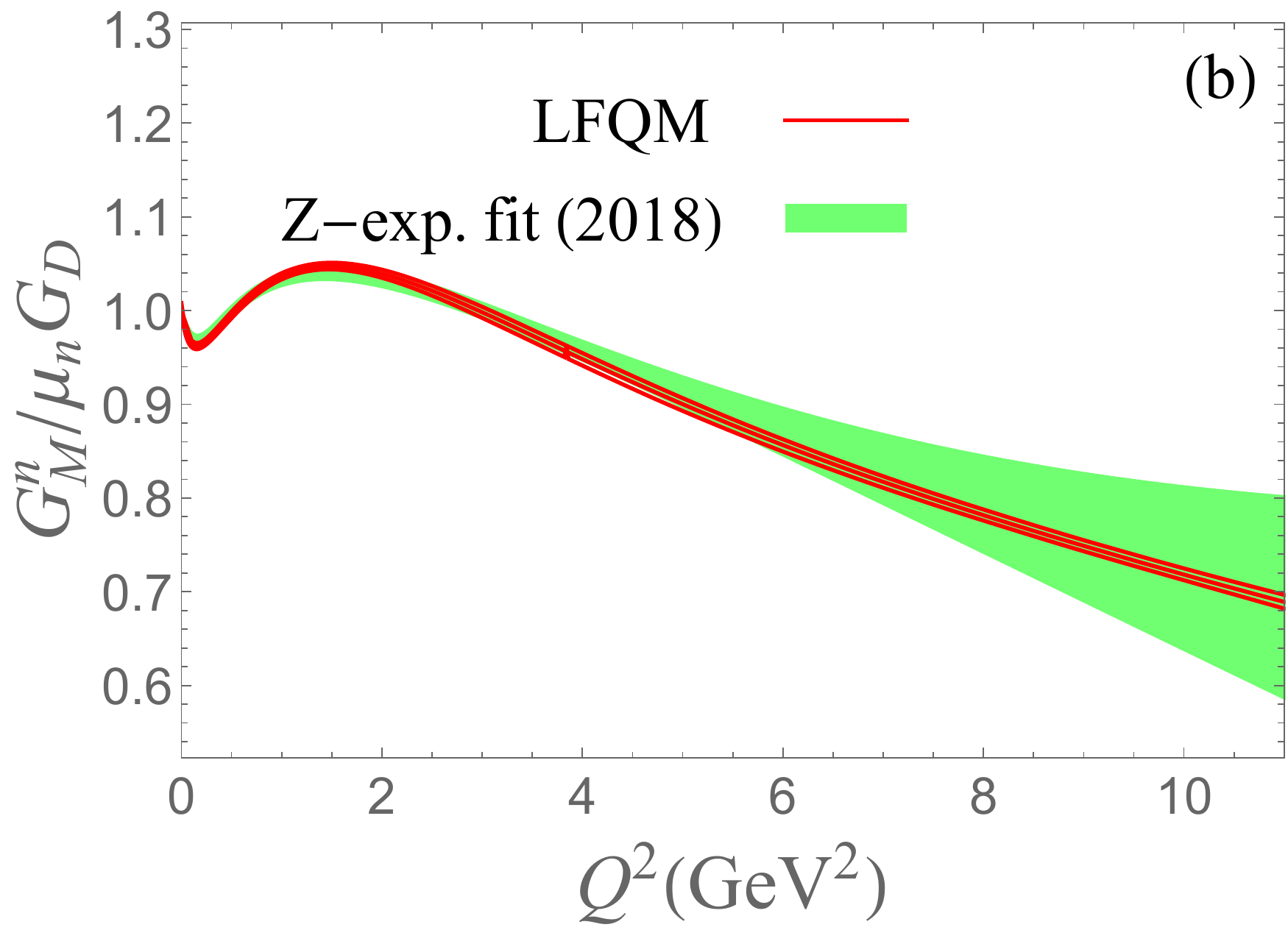}
\caption{\label{fig:Gn} Neutron electric and magnetic form factors. See the caption of Fig.~\ref{fig:Gp} for the illustrations of the legends used here.}
\end{figure}

\begin{table}
 \begin{ruledtabular}
\begin{tabular}{ccccc}
 $c_{11}^s$ & $\beta_{s1}$ & $ c_{20}^s $ & $ c_{21}^s $ & $ \beta_{s2} $     \\ \hline
$0.29_{-1.00}^{+0.67} $ & $0.47_{-0.05}^{+0.05} $ & $ -0.32_{-0.07}^{+0.06} $ & $-3.5_{-0.4}^{+0.5} $ & $ 0.352_{-0.007}^{+0.008} $   \\ \hline \hline 
 $ c_{11}^a $ & $ \beta_{a1} $ & $ c_{20}^a $ & $ c_{21}^a $ & $ \beta_{a2} $  \\ \hline
$ 0.072_{-0.32}^{+0.24} $ & $ 0.52_{-0.04}^{+0.03} $ & $ 6.4_{-1.7}^{+1.6} $ & $  0.5_{-2.6}^{+2.2} $ & $ 0.51_{-0.05}^{+0.04} $  \\ \hline \hline 
$ \mq $ & $ \ms $ & $ \ma  $ & $ \Lambda_N $ & $ \Lambda_\Delta$  \\ \hline 
$ 0.32_{-0.01}^{+0.01 } $ & $ 0.14_{-0.02}^{+0.02} $ & $ 0.35_{-0.05}^{+0.03} $ &  $0.49_{-0.04}^{+0.03}  $ & $ 0.43_{-0.02}^{+ 0.02}$  \\
\end{tabular} \caption{ Parameter mean values and their error bars corresponding to $68\%$ degree of belief.} \label{tab:paralist}
\end{ruledtabular}
\end{table}

To calibrate our model, we rely on a recent analysis of the nucleon's elastic EM form factors in Ref.~\cite{Ye:2017gyb}. The study applied the $z$-expansion approach to parametrize the form factors' $Q^2$ dependence with minimal model assumptions, and then fitted them to the existing measurements.  
The predicted form factors and their error bars are used as ``data'' to constrain the aforementioned model parameters. Specifically, we pick 16 different $Q^2$ values %%GM
 for each of nucleon's four EM form factors, 
\begin{eqnarray}
G_{{E p,n}}(Q^2) &\equiv & F_{{1p,n}} - \tau  F_{{2p,n}}  \\ 
G_{{M p,n}}(Q^2) &\equiv & F_{{1p,n}} +  F_{{2p,n}}  \ . 
\end{eqnarray}
Eight of them are evenly distributed in the $0.01 \leq Q^2 \leq 1.5\, \mathrm{GeV}^2$ region, with the other eight also evenly distributed in  
$ 1.5 < Q^2 \leq 10\, \mathrm{GeV}^2$. 

\begin{figure}
\includegraphics[width=0.4 \textwidth, angle=0]{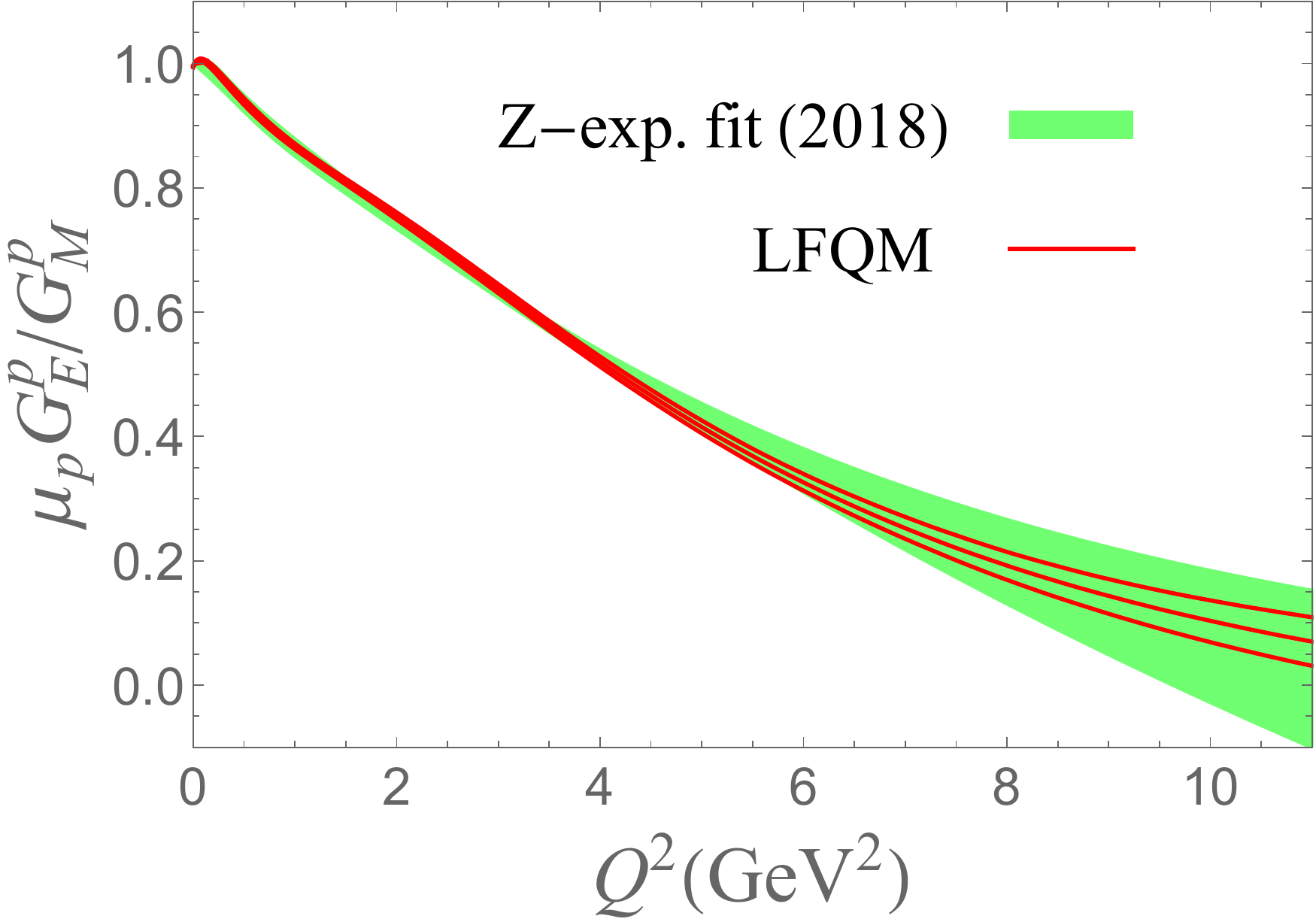} 
\caption{\label{fig:GepOverGmp} The ratio between proton's electric and magnetic form factors. The legend is the same as that in Fig.~\ref{fig:Gp}. }
\end{figure}

The Bayesian inference~\cite{SiviaBayesian96} is then used to compute the posterior probability distribution function (PDF) of the unknown parameter vector, schematically labeled as vector $\vec{g}$, given the existing ``data'' $D$, our theory $T$, and prior information $I$. According to the Bayes' theorem~\cite{SiviaBayesian96}, the desired PDF is related to the likelihood function through 
\begin{equation}
{\rm pr} \left(\vec{g}\vert D;T; I \right)  
=
{\rm pr} \left(D \vert \vec{g};T; I \right) {\rm pr} \left(\vec{g} \vert I \right) . \label{eqn:bayesian1}
\end{equation}
The first term on the right side is proportional to the likelihood:
\begin{equation}
\ln {\rm pr} \left(D \vert \vec{g};T;I \right)  =  c - \sum_{j=1}^N \frac{\left[ F(\vec{g}; Q^2_j)-D_j\right]^2}{2  \sigma_j^{2}},
\end{equation}
where $F(\vec{g}; Q^2_j)$ is the form factor prediction at $Q^2_j$ of the $j$th data point $D_j$, and $\sigma_j$ is the statistical uncertainty associated with $D_j$.  The constant $c$ ensures ${\rm pr} \left(\vec{g} \vert D;T; I \right)$ at the right side is properly normalized. The second term in the right side of Eq.~(\ref{eqn:bayesian1}), ${\rm pr}\left( \vec{g}\vert I \right)$, is the prior for all the parameters $\vec{g}$. It is separable: the priors for $c^{s,a}_{1,1}$, $c^{s,a}_{2,0}$, and $c^{s,a}_{2,1}$ are Gaussian distributions centered at $0$ and with width equal to $5$, while the priors for other parameters are uniform distributions requiring all the $\beta_{}$s between $0$ and $2$~GeV, $ 0.1 \leq \mq \leq 0.6 $ GeV, $\ms$ and $\ma$ between $0$ and $\mn$, $\Lambda_N$ and $\Lambda_\Delta$ between $0$ and $1.5$~GeV. It should be pointed out that in this work, the errors of ``data'' at our picked $Q^2$ values are treated as uncorrelated, considering the correlation information for the ``data'' are not available in public. This simplification needs to be further improved in %%GM
a  future study. 

The Markov Chain Monte Carlo method is then employed to sample the posterior PDF in the 15 dimension space. The particular sampling algorithm is the so-called {\tt emcee} sampler~\cite{2013PASP..125..306F} coupled with parallel tempering~\cite{2016MNRAS.455.1919V}. The sampler has been extensively used in {\it e.g.}, astronomy for the same purpose~\cite{2013PASP..125..306F, 2016MNRAS.455.1919V}. The detailed 2-dim and 1-dim projections of this PDF can be seen in  Fig.~\ref{fig:corrplt}. The central values and $68\%$ degrees-of-belief error bars of the model parameters can be found in Table~\ref{tab:paralist}. The $c^{s,a}_{1,1}$, $c^{s,a}_{2,0}$, and $c^{s,a}_{2,1}$ parameters are constrained to regions with widths at most about half of the Gaussian prior widths, while the other parameters are very localized within the windows of their uniform priors. Therefore, enlarging the prior windows for these parameters will not significantly modify the posterior PDF, ${\rm pr} \left(\vec{g}\vert D;T; I \right)$. It is also interesting to note that the preferred parameter values  are consistent with the naive expectation raised in the previous section.

With the samples of the posterior PDF, we can compute the central value and error bar for any quantity as  a function of $\vec{g}$. Figs.~\ref{fig:Gp}, \ref{fig:Gn}, and~\ref{fig:GepOverGmp} plot our error bands (the red curves) for the nucleon EM form factors---normalized against the $G_D(Q^2)$---and proton's form factor ratio, to be compared with the results (the green bands) from Ref.~\cite{Ye:2017gyb}. Note the normalizations for magnetic form factors $\mu_p = 2.793$ and $\mu_n = -1.913$ are from the supplementary material of Ref.~\cite{Ye:2017gyb}. The model results are in good agreement with the ``data''. In particular, the $G_E^p$-$G_M^p$ ratio as shown in Fig.~\ref{fig:GepOverGmp} agrees very well with the extraction from Ref.~\cite{Ye:2017gyb} in the shown $Q^2$ window, which is an improvement over the previous calculations using similar approach~\cite{Miller:2002ig,Matevosyan:2005bp}.  However, the difference between our $G_{E}^n$ result and the ``data'', as shown in Fig.~\ref{fig:Gn}, shows that our model prefer smaller values for $G_{E}^n$ at momentum transfer above $4~\mathrm{GeV}^2$. Moreover, our error bars are consistently smaller than those from Ref.~\cite{Ye:2017gyb}. Possible reasons include missing correlation between ``data'' in our inference, and/or the absence of theoretical uncertainty of our quark-diquark model.

\begin{figure}
\includegraphics[width=0.4 \textwidth, angle=0]{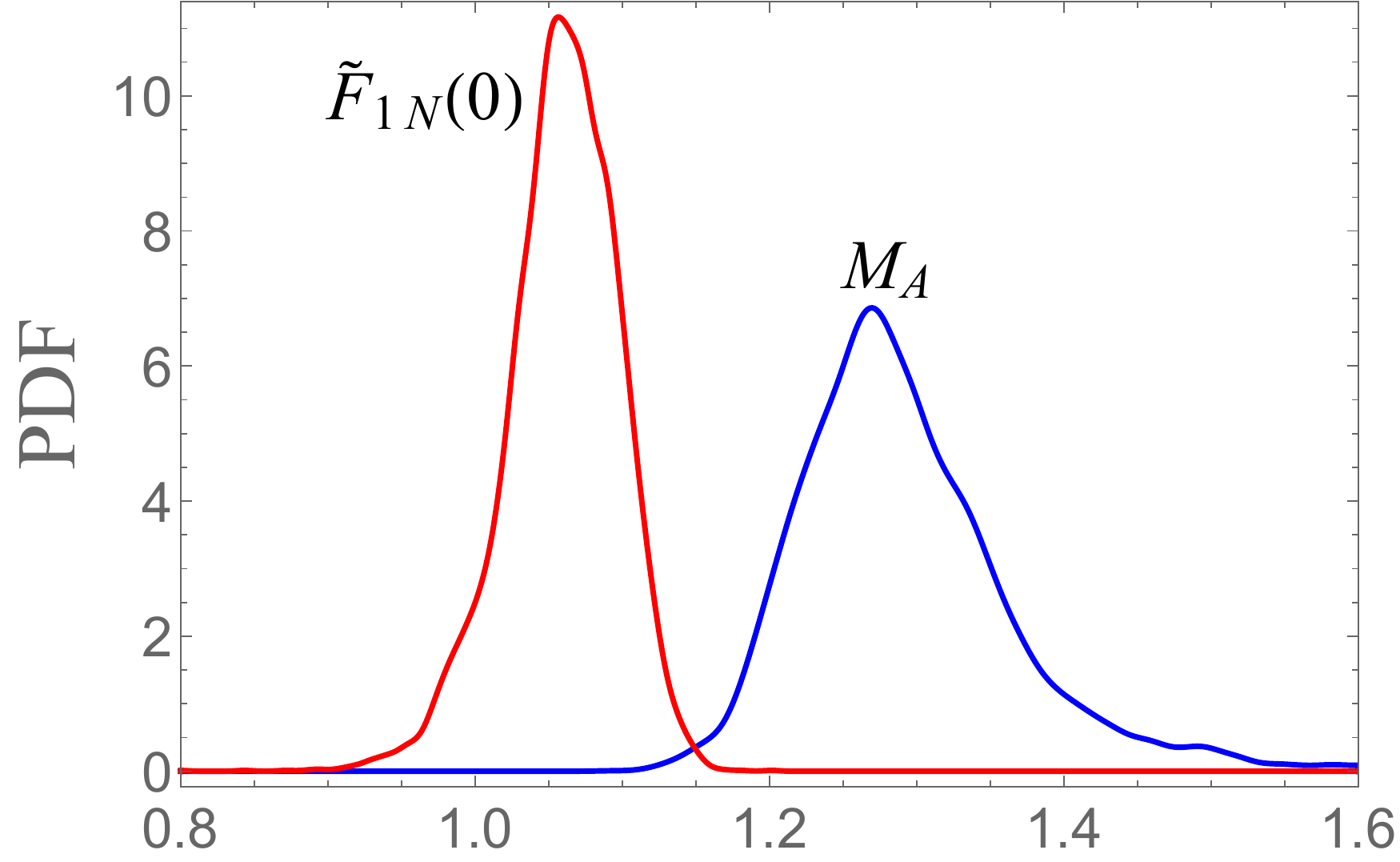} 
\caption{\label{fig:FA0}1-dim PDFs for $M_A$ (blue curve in the unit of GeV) and $\FAn{1}(0)$ (red curve) from our Bayesian inference.}
\end{figure}

Turning to the axial form factor $\FAn{1}$: the 1-dim posterior PDFs for $\FAn{1}(Q^2=0)$ and the $M_A$ value extracted from the first derivative of $\FAn{1}$ at $Q^2=0$ are plotted in
Fig.~\ref{fig:FA0}. Our prediction for $\FAn{1}(Q^2=0)$ is $1.06 \pm 0.04$, which is somewhat smaller than $g_A = 1.27$; $r_A^2 = 0.29 \pm 0.03\, \mathrm{fm}^2$ and the associated
$M_A= 1.28 \pm 0.07$ GeV. The $r_A^2$ is smaller than $r_A^2 = 0.46 \pm 0.16\, \mathrm{fm}^2$ from a recent analysis~\cite{Hill:2017wgb} (the associated $M_A= 1.01 \pm 0.17\, \mathrm{GeV}^2$)
based on existing neutrino-nucleon scattering and muon weak capture data, and closer to current Lattice QCD results having $r_A^2$ ranging from $0.2$
to $0.45~\mathrm{fm}^2$. Although our $r_A^2$ is within the 1-$\sigma$ band of the recent analysis~\cite{Hill:2017wgb}, the uncertainty assigned for our $r_A^2$ prediction  is too small to cover the latter's central value. However our error bar only accounts for that within our model parameter space, while the theoretical uncertainty of the current model is difficult to estimate and not included in the error bar. 

\begin{figure}
\includegraphics[width=0.4 \textwidth, angle=0]{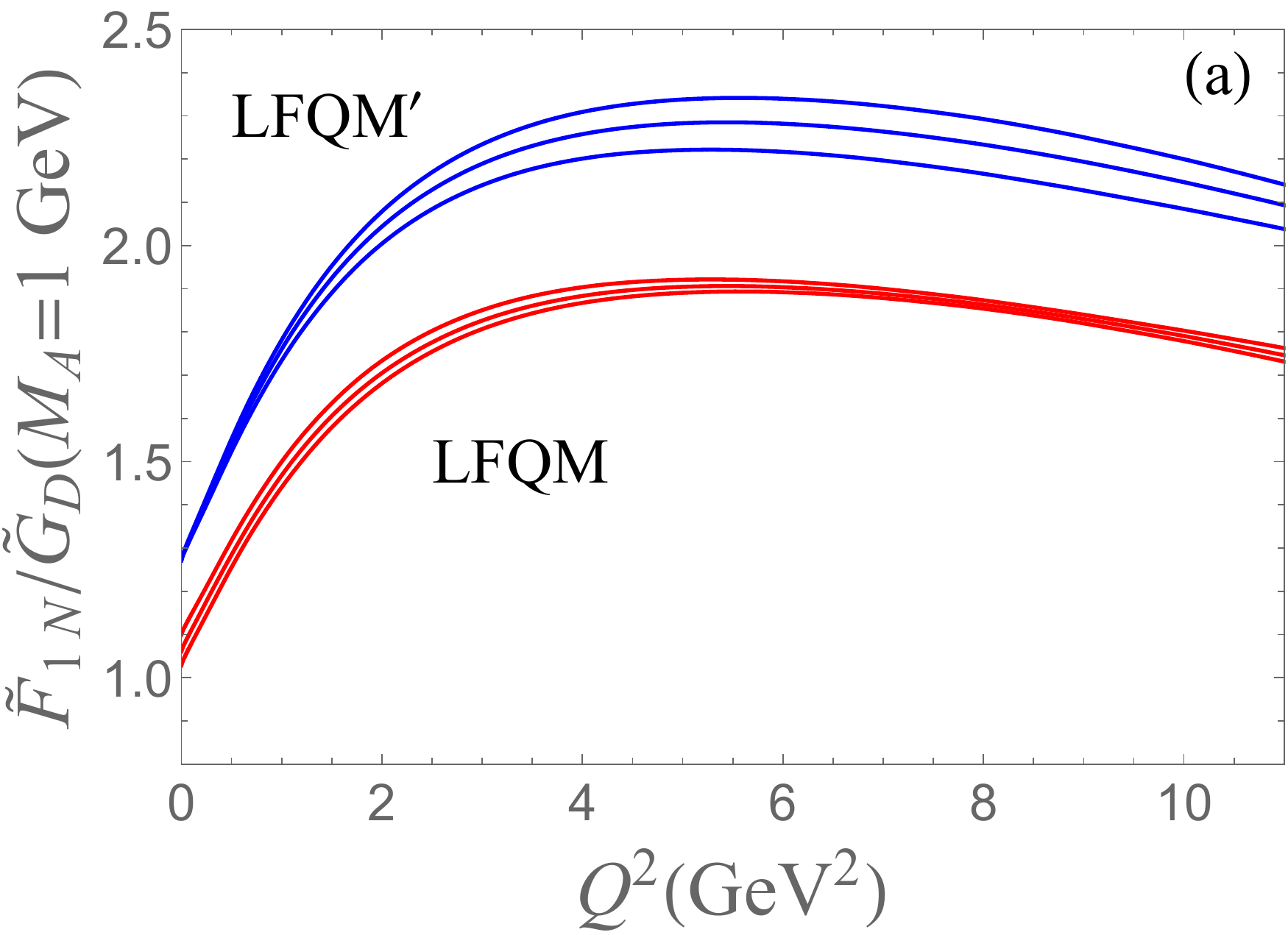} 
\includegraphics[width=0.4 \textwidth, angle=0]{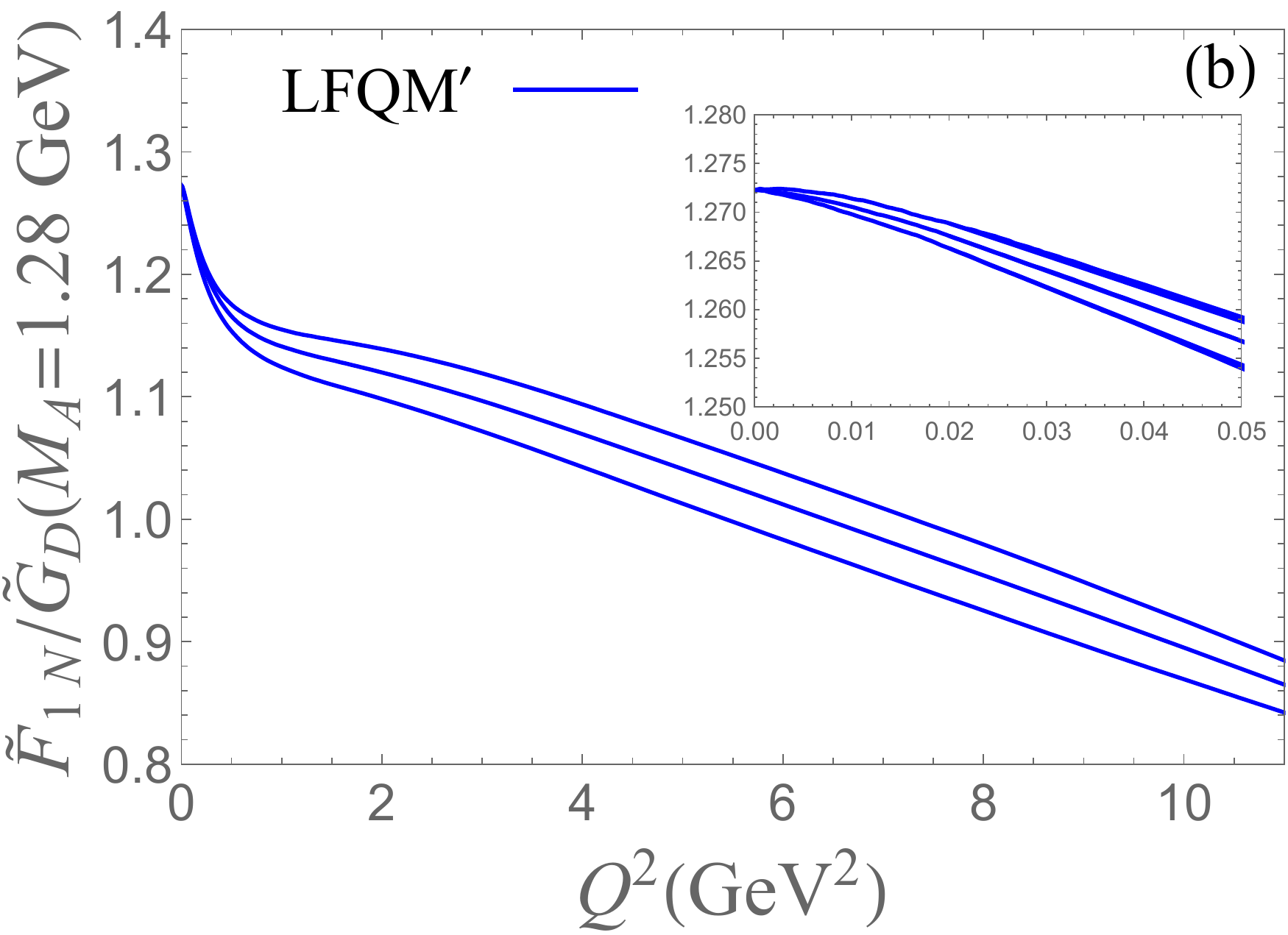}
\caption{\label{fig:FA} The model prediction of $\FAn{1}$ and its error band. Panel (a) normalizes form factor against $\tilde{G}_D$ with $M_A=1$ GeV, while panel (b) uses $M_A= 1.28$ MeV. Panel (a) shows two sets of curves: ``LFQM'' (red curves) are the model's original prediction, and  the ``$\mathrm{LFQM}'$'' (blue curves) re-scale the curves such that the $Q^2=0$ value agrees with $g_A=1.27$. The inset in panel (b) demonstrates that when $Q^2 \sim 0$, the full form factor is close to the corresponding dipole parametrization.}
\end{figure}

The $\FAn{1}(Q^2)$'s central value and its 1-$\sigma$ lower and upper bounds are shown in Fig.~\ref{fig:FA}, re-scaled by $\tilde{G}_D \equiv (1+ Q^2/M_A^2)^{-2}$ with $M_A=1$ GeV (panel (a)) and $M_A=1.28$ GeV (panel (b)). The latter $M_A$-value is the central value of our analysis. Panel (a) shows two sets of curves: the ``LFQM'' (red curves) are our predictions while each of the ``$\mathrm{LFQM}'$'' (blue curves) re-scale the corresponding ``LFQM'' curves by a constant such that the $Q^2=0$ value agrees with $g_A=1.27$. 
The global rescaling is equivalent to treating the size of quark axial charge $e_{Aq}$ as a fitting parameter, because the contributions from  both bare quark and pion cloud originate from quark's axial charge. (As pointed out in Sec.~\ref{subsec:inputpioncloud}, the contributions from diagram \III~ with $\Delta$ in the loop are normalized by $e_{Aq}$, although they are approximated by the physical inelastic form factors.) 
In the following calculations of cross sections, the rescaled axial form factors are always implicitly assumed.
In panel (b), only the corresponding ``$\mathrm{LFQM}'$'' results  are plotted. We do see a significantly different $Q^2$ dependence from $\tilde{G}_D$ with $M_A= 1 $ GeV; and more importantly that our $\FAn{1}$ differ from its dipole approximation by about $10\%$ at $Q^2 $ between 1 and 2~$\mathrm{GeV}^2$. The latter suggests the necessity of using the full form factor instead of a simple dipole approximation for modeling neutrino-nucleus QE scatterings in the coming neutrino-oscillation experiments. 

\begin{figure}
\includegraphics[width=0.4 \textwidth, angle=0]{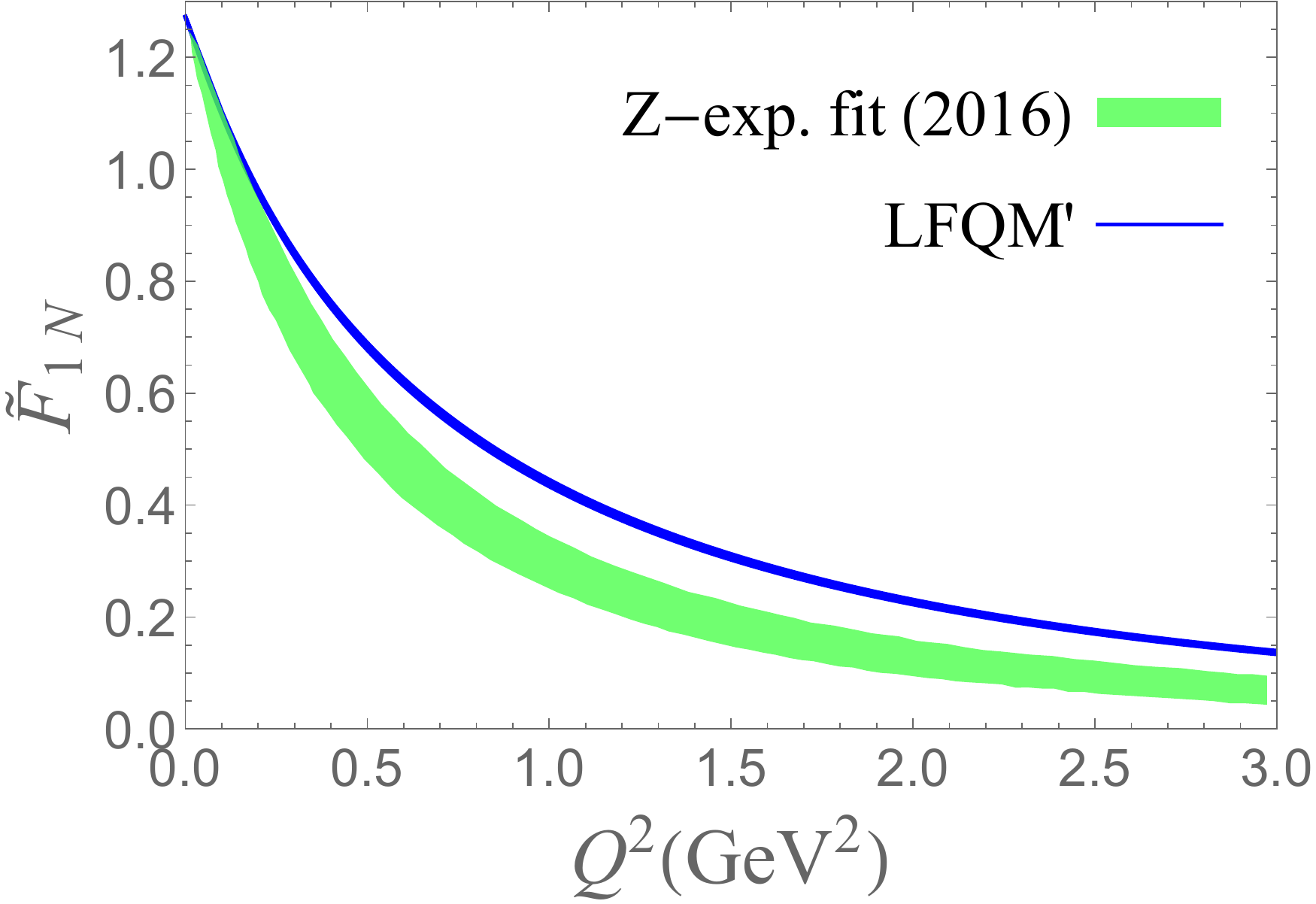} 
\caption{\label{fig:FAvsHill}Comparison between $\mathrm{LFQM}'$ $\FAn{1}(Q^2)$ and the same form factor fitted based on the $z$-expansion approach of
Ref.~\cite{Meyer:2016oeg}.}
\end{figure}

Fig.~\ref{fig:FAvsHill} compares our $\FAn{1}$ with the $z$-expansion-based fit from Ref.~\cite{Meyer:2016oeg}. The comparison in the low-$Q^2$ region is consistent with
the discussion above concerning $r_A^2$, whereas for $Q^2$ beyond 0.5~$\mathrm{GeV}^2$, our form factor becomes increasingly larger than that obtained using the $z$-expansion
of Ref.~\cite{Meyer:2016oeg}. As a compelling extension of this work, it would be interesting to investigate the theoretical description of previous neutrino-deuteron scattering
measurements (cf.~Refs.~\cite{Meyer:2016oeg, Hill:2017wgb}), by combining the LFQM calculations in this analysis with a systematic treatment of deuteron-structure effects and
the subtleties associated with these experiments~\cite{Meyer:2016oeg}. We reserve such an undertaking to future efforts.

We can also parameterize our $\FAn{1}(Q^2)/\FAn{1}(0)$ by employing the $z$-expansion form from Ref.~\cite{Meyer:2016oeg}.  With $t_\mathrm{cut}= 9 m_\pi^2$, $t_0 = - 1.19263\,\mathrm{GeV}^2$, and 
\begin{eqnarray}
z(Q^2) & \equiv & \frac{\sqrt{t_\mathrm{cut} + Q^2} - \sqrt{t_\mathrm{cut} - t_0}}{\sqrt{t_\mathrm{cut} + Q^2} + \sqrt{t_\mathrm{cut} - t_0}}  \ ,   
\end{eqnarray} 
the central value of our axial form factor can be parametrized as 
\begin{eqnarray}
\frac{\FAn{1}(Q^2)}{\FAn{1}(0)} &=& \sum_{k=0}^{11} a_k z^k(Q^2) \ .  \label{eqn:zexpforaxial}
\end{eqnarray}
This parametrization reproduces the full $\FAn{1}(Q^2)$ with better than $0.1\%$ error for $Q^2$ between 0 and 10~$\mathrm{GeV}^2$. The numerical values of the  $a_k$ coefficients can be found in Table~\ref{tab:ak}.

\begin{table}
 \begin{ruledtabular}
\begin{tabular}{cccccc}
 $k=0$ & $1$ & $ 2$ & $ 3 $ & $4 $ & $5 $     \\ \hline
$0.299145$ & $-1.18966 $ &  $1.16692$ & $0.763023$ & $-0.39146$ & $-2.45022$    \\ \hline \hline 
$6$ & $7$ & $ 8$ & $ 9 $ & $ 10 $ & $ 11 $       \\ \hline
 $-8.74781 $  & $ 23.8158 $ &  $48.8291$ & $-126.237$ & $-103.061$ & $259.714$   \\
\end{tabular} \caption{The fitted values for $a_k$ as used in the $z$-parametrization in Eq.~(\ref{eqn:zexpforaxial}) for the central value of ${\FAn{1}(Q^2)}/{\FAn{1}(0)}$.} \label{tab:ak}
\end{ruledtabular}
\end{table}

\section{Impacts} \label{sec:Impacts}
In order to quantify the impact of the difference between our full $\FAn{1}$ and the commonly used dipole approximation, we first calculate the cross sections for the charged-current (CC) (anti)neutrino--nucleon scattering and then the (anti)neutrino--$^{40}\mathrm{Ar}$ QE scatterings relevant for the coming DUNE experiment.  

\subsection{The single-nucleon cross section} 

\begin{figure*}
\includegraphics[width=0.4 \textwidth, angle=0]{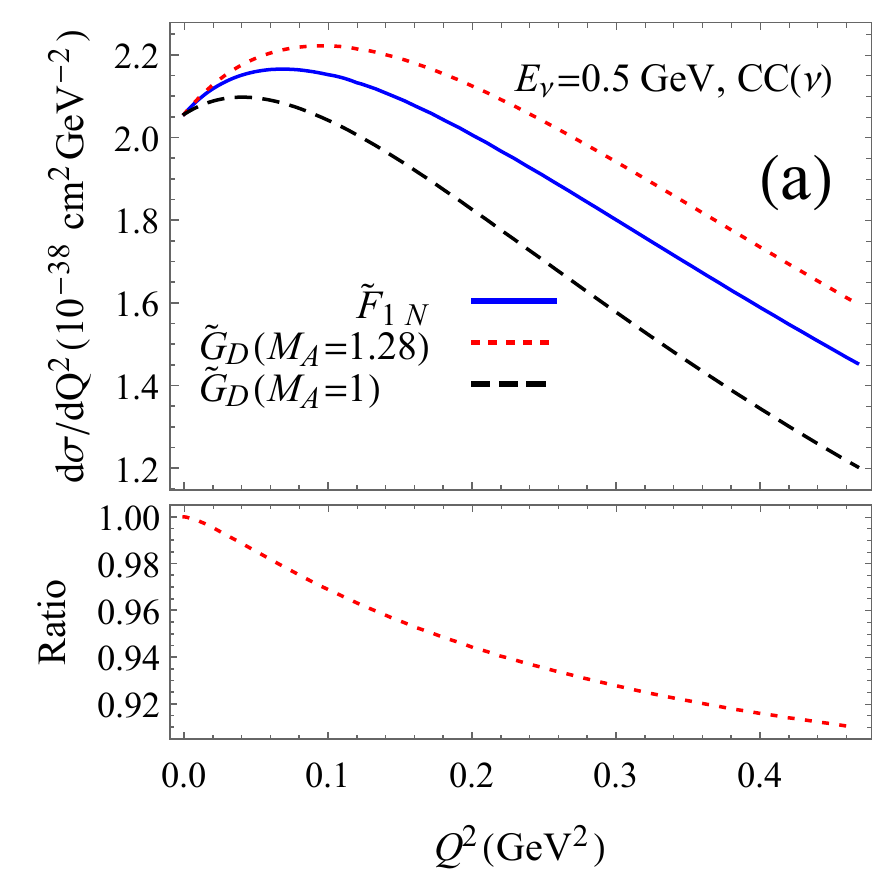} 
\includegraphics[width=0.4 \textwidth, angle=0]{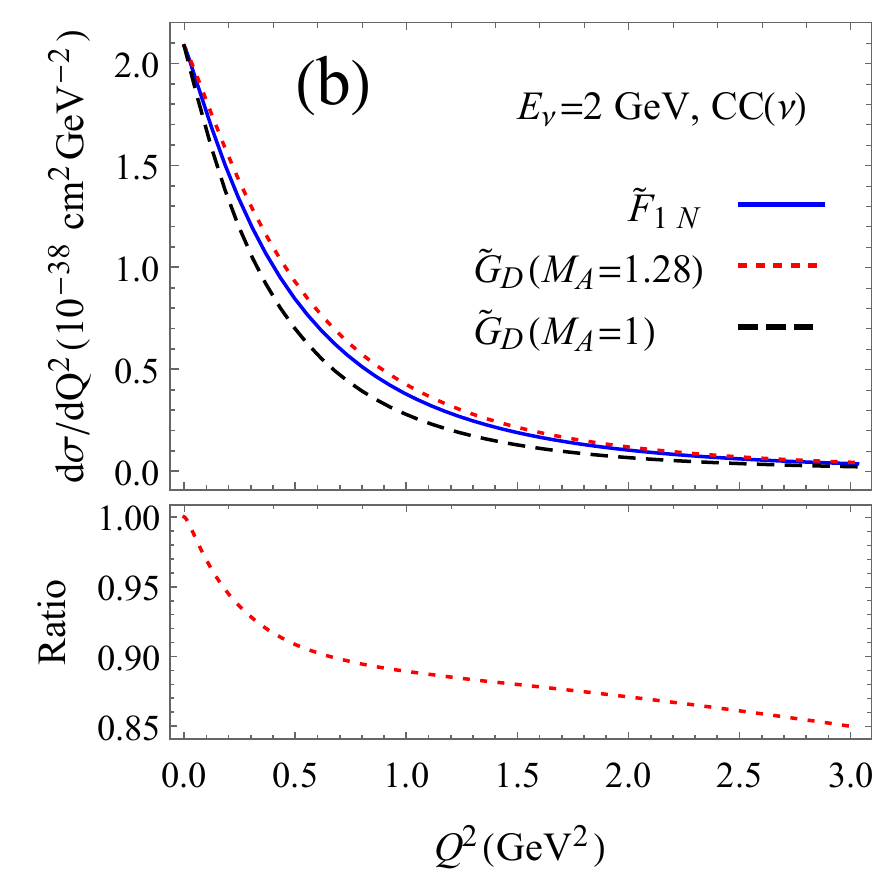}
\caption{\label{fig:dsigdQsq_nu} Differential cross section for neutrino scattering at $E_\nu = 0.5$ and $2$ GeV. In the upper panels, three different calculations are plotted with different axial form factor, while the lower panels show the ratio between the result using our full form factor and the one using $g_A \tilde{G}_D$ with $M_A=1.28$ GeV. }
\end{figure*}

\begin{figure*}
\includegraphics[width=0.4 \textwidth, angle=0]{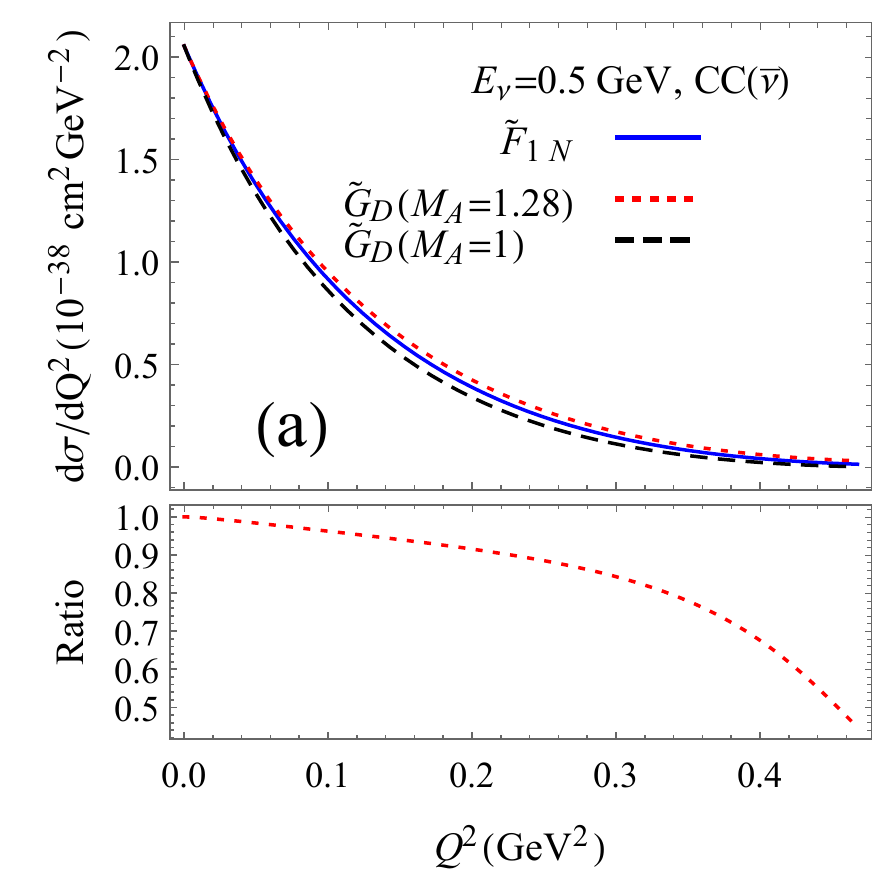} 
\includegraphics[width=0.4 \textwidth, angle=0]{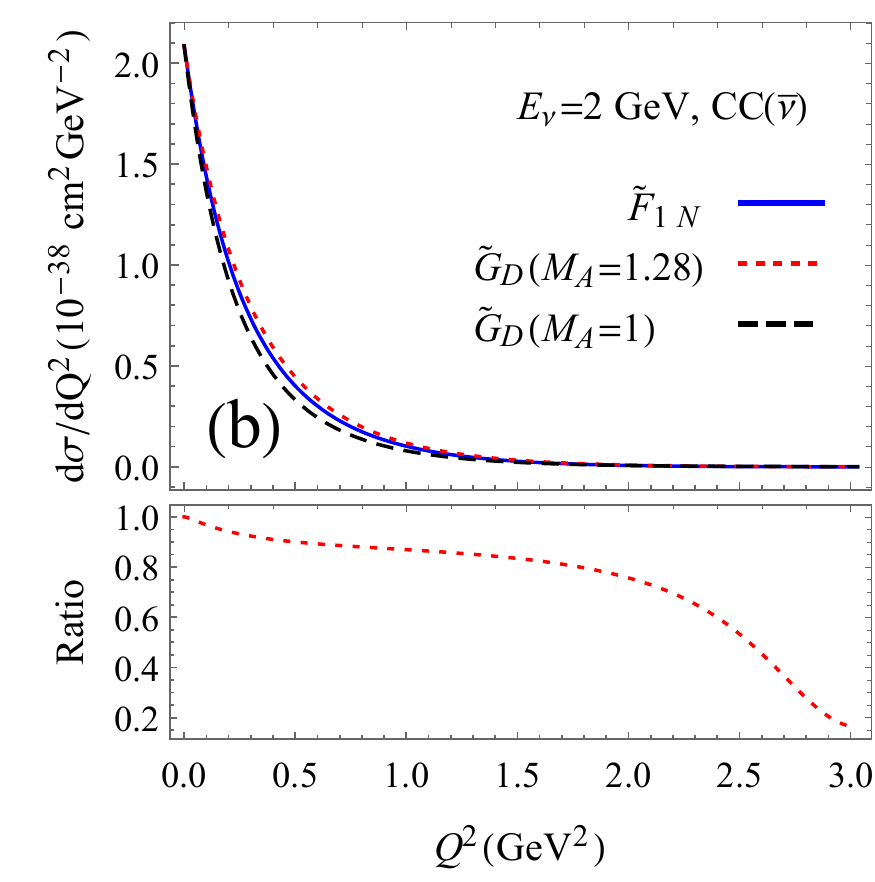}
\caption{\label{fig:dsigdQsq_antinu} Differential cross section for anti-neutrino scattering at $E_\nu = 0.5$ and $2$ GeV. See the caption of Fig.~\ref{fig:dsigdQsq_nu} for the illustrations of the legends.}
\end{figure*}

The single-nucleon scattering cross section differentiated against $Q^2$ at given neutrino energy $E_\nu$ can be written as ~\cite{LlewellynSmith:1971uhs}
\begin{align}
	\frac{d \sigma^{\nu (\bar{\nu})}}{d Q^2} \equiv \frac{G_F^2\cos^2\theta_c \mn^2}{8\pi E_\nu^2} \Big[A &\mp B \frac{s-u}{\mn^2} \\
	&+ C  \frac{\left(s-u\right)^2}{\mn^4} \Big] \notag
\end{align}
with $G_F$ as Fermi constant, $\theta_c$ as Cabibbo angle; $\mmu$ as the charged lepton mass; $s-u = 4 E_\nu \mn - Q^2 -\mmu^2$; the sign of $B$: $(-)$ for the neutrino scattering and  $(+)$ for the antineutrino scattering;  and 
\begin{eqnarray}
A &=&  \frac{\left(\mmu^2+Q^2\right)}{\mn^2} \bigg\{-(1-\tau) F_{1V}^2 +4 \tau F_{1V} F_{2V} \notag \\ 
   && + (1-\tau) \tau F_{2V}^2 + (1+\tau) \FAn{1}^2 \notag \\ 
	&& \hspace*{-0.3cm} -\frac{\mmu^2}{4\mn^2} \left[(F_{1V}+F_{2V})^2+(\FAn{1}+\FAn{2})^2-\FAn{2}^2 (1+\tau)\right] \bigg\}  \notag \\ 
B &=& 4 \tau \FAn{1} \left(F_{1V}+F_{2V} \right)  \notag \\ 
C  &=& \frac{1}{4} (\FAn{1}^2 + F_{1V}^2 + \tau  F_{2V}^2 ) \ . \notag 
\end{eqnarray}
Here, $F_{1V} \equiv F_{1p}-F_{1n}$, $F_{2V} \equiv F_{2p}-F_{2n}$ are the form factor for the isovector component in the EM current. When integrating the differential cross section over the $Q^2$ to get the total cross section, the range of $Q^2$ depends on neutrino Lab energy $E_\nu$; its lower and upper limits are 
\begin{equation} 
\frac{2 E_\nu^2}{\left(1 + 2 \frac{E_\nu}{\mn} \right)} \left[\mathcal{R}+ 1 \mp \sqrt{(\mathcal{R}- 1)^2 - \frac{\mmu^2}{E_\nu^2}}  \right] - \mmu^2 
\end{equation}
with $\mathcal{R}\equiv  \frac{\mmu^2}{2 \mn E_\nu}$. Meanwhile, the threshold for $E_\nu$ is 
\begin{equation}
E_\nu \geq \mmu + \frac{\mmu^2}{2\mn}\ . 
\end{equation}

\begin{figure*}
\includegraphics[width=0.4 \textwidth, angle=0]{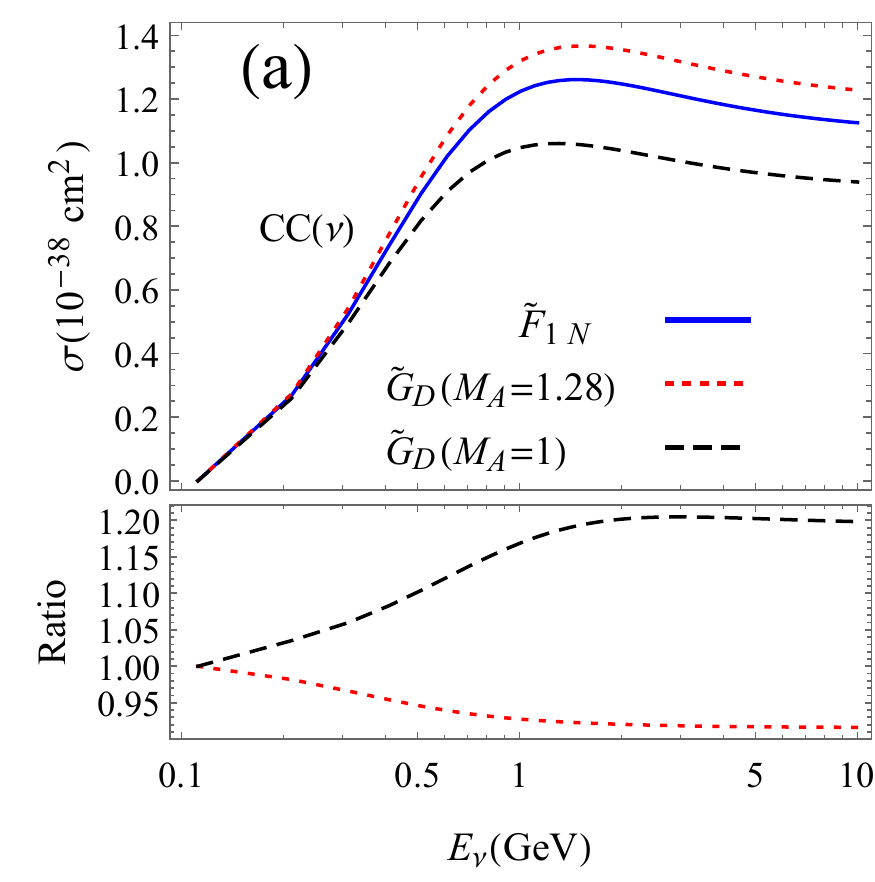} 
\includegraphics[width=0.4 \textwidth, angle=0]{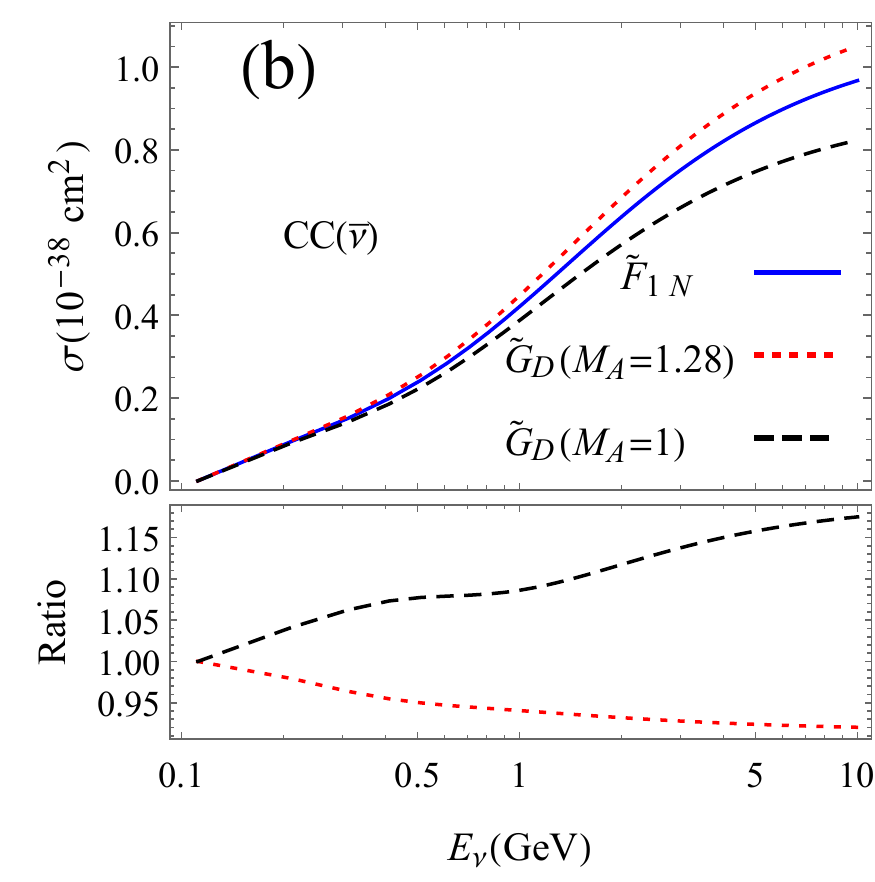}
\caption{\label{fig:sig} Total cross section for CC-induced neutrino and antineutrino scattering off nucleon. In the upper panels, three different calculations are plotted with different axial form factor, while the lower panels show the ratios between the results using the full $\FAn{1}$ and the one using $g_A \tilde{G}_D$ and $M_A=1.28$ GeV (red dotted curve) and with the results using $g_A \tilde{G}_D$ and $M_A=1$ GeV.  }
\end{figure*}

Figs.~\ref{fig:dsigdQsq_nu} and~\ref{fig:dsigdQsq_antinu} compares differential cross section due to three different axial form factor in the CC-induced
(anti)neutrino scatterings. Two different $E_\nu = 0.5$ and $2$ GeV are chosen. We see even having $M_A=1.28$ GeV such that the dipole parametrization
agrees with the full form factor at $Q^2 \sim 0 $, their cross section results can differ up to $5$-$10\%$ in the dominating $Q^2$ regions.
Fig.~\ref{fig:sig} shows the total cross sections vs $E_\nu$ based on those form factors: the difference between the full-form-factor based
calculations and the dipole-parametrization based ($M_A=1.28$ GeV) increases to around $5\%$ at about $E_\nu \sim 0.5$ GeV and mildly increase
to a little below $8\%$ with $E_\nu =10$ GeV. This $E_\nu$ range covers the dominating region of the DUNE's neutrino spectra.
As such, we conclude that controlling these effects within the QE cross section will be critically important to further strengthening the interpretation
of results from the upcoming DUNE program. Of course, the difference between the full calculation and the $M_A =1 $ GeV one is much larger than the previous ones, reaching to $20\%$ above 1  GeV neutrino energy. Note in all the figures, the EM form factors are the full form factor from our model.

\subsection{Neutrino-nucleus cross sections}
\begin{figure}
\includegraphics[width=0.4 \textwidth, angle=0]{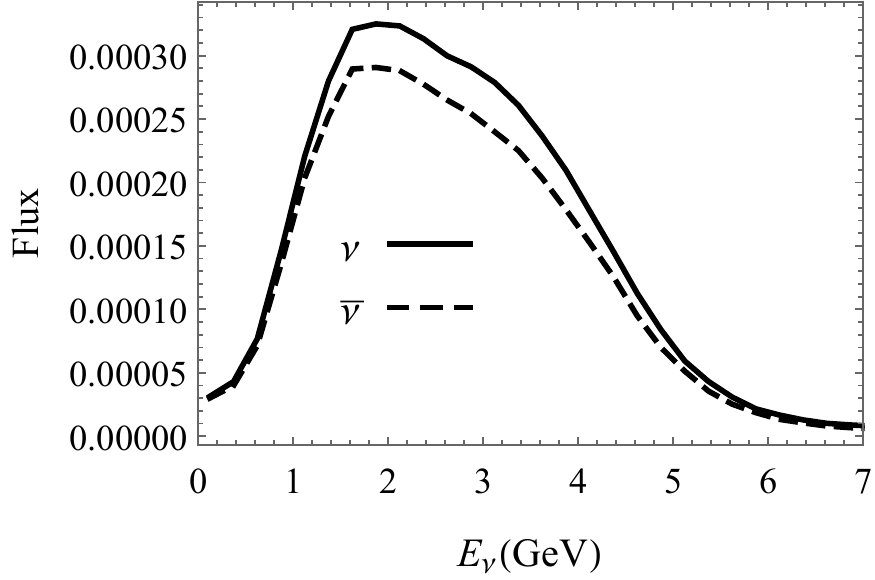}
\caption{\label{fig:LBNEflux}. The $\nu_\mu$ and $\bar{\nu}_\mu$ fluxes in neutrino and antineutrino mode in the DUNE experiment's near detector ~\cite{Acciarri:2015uup}. The units are irrelevant in this work.  }
\end{figure}

To study the form factor's impact on the neutrino-nucleus cross sections relevant for the DUNE experiment~\cite{Acciarri:2015uup}, we use the GiBUU package to compute the $\nu$($\bar{\nu}$)--$^{40}\mathrm{Ar}$ \emph{inclusive} QE scattering \cite{Buss:2011mx,Mosel:2016cwa}. The initial state nuclear effects, including Fermi motion, are automatically taken into account, while the final state interaction is not relevant and thus turned off in the simulations.  The two-particle-two-hole process, resonance and pion productions, and deep inelastic scatterings are not studied here. The neutrino fluxes (see Fig.~\ref{fig:LBNEflux}) in our calculation are the so-called ``Reference,  204x4 m DP'' from Ref.~\cite{Acciarri:2015uup}. Note that in the calculations here, we simply use the vector current form factors~\cite{Bodek:2007ym} native to the GiBUU package.

\begin{figure*}
\includegraphics[width=0.4 \textwidth, angle=0]{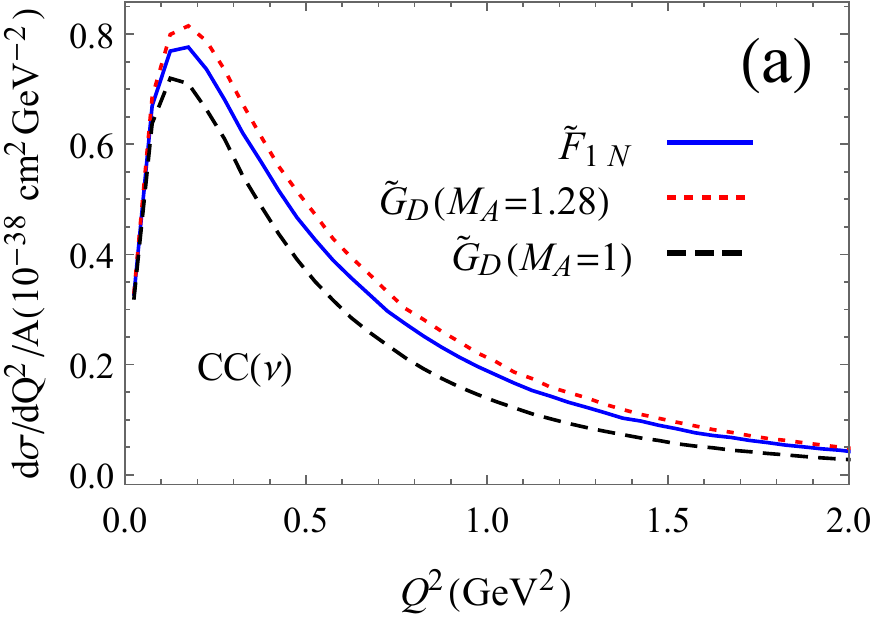} 
\includegraphics[width=0.4 \textwidth, angle=0]{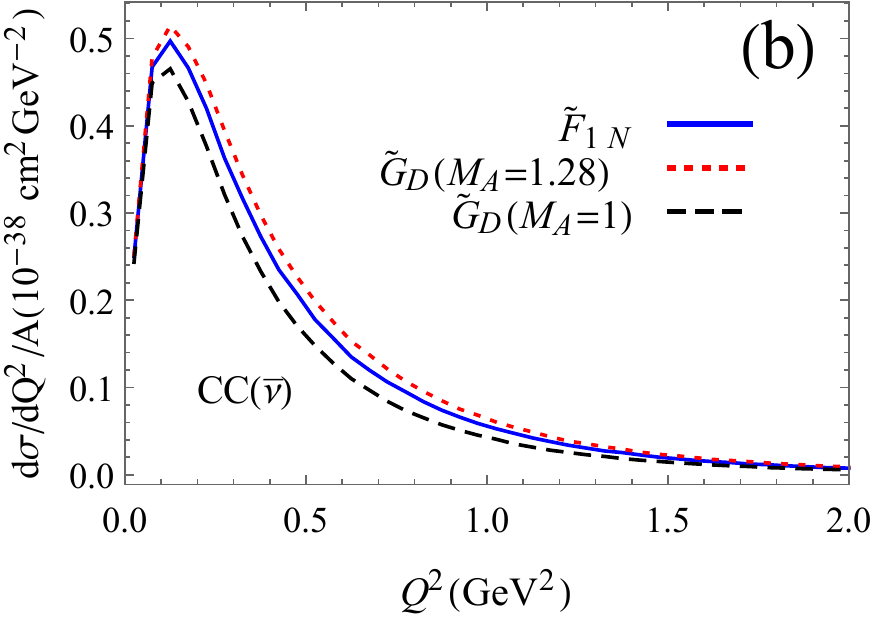}
\includegraphics[width=0.4 \textwidth, angle=0]{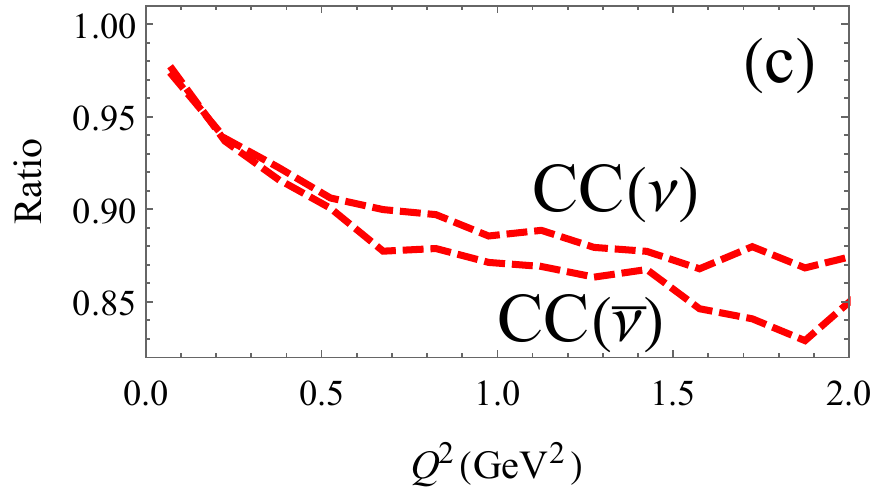}
\caption{\label{fig:dsigdQsqLBNE} The DUNE-flux averaged $\nu$($\bar{\nu}$)--$^{40}\mathrm{Ar}$ scattering differential cross section. The lower panel again shows the ratio between the result using our model form factor and the one using its dipole approximation with $M_A=1.28$ GeV. }
\end{figure*}

Panel (a) and (b) in Fig.~\ref{fig:dsigdQsqLBNE} show the DUNE flux-averaged differential cross section vs $Q^2$ for both neutrino and antineutrino scatterings. Panel (c) shows the ratios between the full-axial-form-factor based and the dipole-parametrization-based (with $M_A=1.28$ GeV) calculations.  Indeed the difference is about $5\%$ in the dominant $Q^2$ region around $0.2$~$\mathrm{GeV}^2$ and increases to about $10\%$ at $Q^2 \sim 1$~$\mathrm{GeV}^2$ and beyond. The wiggles in the tails of the ratio plot is due to the diminishing simulation statistics in the large $Q^2$ region. It is worth noting that, in panel (c), for $Q^2$ below $0.5\,\mathrm{GeV}^2$, the differences between the two calculations in both neutrino and antineutrino scatterings are almost the same, but then differ at a few percent level with $Q^2$ a little above $0.5\,\mathrm{GeV}^2$.

\section{Summary}
\label{sec:conc}
In this work, the light-front quark model with pion cloud is employed to correlate the nucleon's EM form factors with its axial form factors. The model is calibrated to the EM form factors' measurements, and then used to predict the axial form factor $\FAn{1}$. We found our form factor's  $r_A^2 = 0.29 \pm 0.03 \,\mathrm{fm}^2$; its central value is smaller than the one resulted from a recent analysis~\cite{Hill:2017wgb} neutrino-nucleon scattering data and the singlet muonic hydrogen capture rate measurement, $r_A^2 = 0.46 \pm 0.16 \,\mathrm{fm}^2$, although the former's central value is within the latter's 1-$\sigma$ error bar. Meanwhile, our value is closer to the current Lattice QCD results from $0.2-0.45\, \mathrm{fm}^2$, although these Lattice calculations still have room to be improved~\cite{Hill:2017wgb}. Note the corresponding $M_A = 1.28 \pm 0.07$ MeV (based on the form factor's $Q^2$ derivative at zero) is larger than $M_A =1.01 \pm 0.17$~$\mathrm{GeV}^2$ from Ref.~\cite{Meyer:2016oeg,Hill:2017wgb}).

More importantly, we found the widely used dipole approximation to our full $\FAn{1}$ over-estimates the (anti)neutrino scattering cross sections, as compared to the calculation using the full expression, by  about $5\%$ at neutrino energy around $0.5$ GeV and reaches about $10\%$ at 10 GeV. By using the GiBUU simulation package, we studied how this discrepancy could impact the (anti)neutrino \emph{inclusive} QE cross sections (without short-range-correlation's contribution and pion production mechanisms) for the coming DUNE experiment. The flux-averaged differential cross section vs $Q^2$ could get over-estimated by $5\%$ at the $Q^2 \sim 0.2$~$\mathrm{GeV}^2$ and reaches around $10\%$ at $1$~$\mathrm{GeV}^2$. The difference between the over-estimation in the neutrino scattering and that in the antineutrino scattering increases to a few percent level when $Q^2$ goes above $0.5$~$\mathrm{GeV}^2$, which could be relevant for the neutrino-oscillation measurements interested in the difference between neutrino and antineutrino. 

In the current work,  
fitting model parameters was simplified by using 
the results of the data analysis (based on the $z$-expansion) from Ref.~\cite{Ye:2017gyb}. For simplicity, the correlations between their extracted form factors---not available in public---are ignored in our model calibration. In the future, our model calibration can be improved either by  directly using the experimental cross section measurements or by including the correlations among the form-factor errors at different $Q^2$ from Ref.~\cite{Ye:2017gyb}. Moreover, the theoretical uncertainty of our quark model is not fully explored, even though we have used a somewhat flexible parametrization of the quark-diquark wave functions. 

In the pion-cloud calculation, the $\Delta$ resonance's contribution is computed by using its form factors either from Lattice QCD calculations or experimental measurements. However a more consistent approach 
would be  to base the inelastic form factor used in the pion-cloud calculations on the $\Delta$'s light-front wave functions. This will also allow studying the axial inelastic form factors, which are also poorly constrained but important for understanding the pion productions in the coming neutrino experiments, within the same framework.

\begin{acknowledgments}
X.Z.~would like to thank Ulrich Mosel for his help with running the GiBUU package.      
We are also thankful to I.~Clo\"{e}t for useful discussions.
X.Z.~was supported by the US Department of Energy under contract DE-FG02-97ER-41014, the US Institute for Nuclear Theory, and the National Science Foundation under
Grant No.~PHY-1614460 and the NUCLEI SciDAC Collaboration under US Department of Energy MSU subcontract RC107839-OSU. T.H.~was supported by the US Department of Energy
under contract DE-SC0010129, and also acknowledges support from a JLab EIC Center Fellowship. G.M.~was supported by the US Department of Energy under contract
DE-FG02-97ER-41014.
\end{acknowledgments}

\appendix

\section{quark wave functions}\label{app:quarkwf}
The scalar-diquark wave functions have already been computed in Ref.~\cite{Cloet:2012cy}, but we present them here for the sake of completeness. We note that the notation used here is somewhat different
from that in Ref.~\cite{Cloet:2012cy}. Our choices for the metric and Dirac spinors follow the Lepage-Brodsky conventions in Ref.~\cite{Brodsky:1997de}. The expression for the bare-quark form factors
(cf.~Eqs.~(\ref{eqn:f1sdef}), (\ref{eqn:f2sdef}), and~(\ref{eqn:f1stildedef})) in terms of the quark light-front wave function, are 

 \begin{table}
 \begin{ruledtabular}
\begin{tabular}{ccc}
 &\multicolumn{2}{c}{$\lambda_{_N}$} \\
$\lambda_q$ & $-\half$ &  $ \half $ \\  \hline
$ -\half $ &  $\varphi_1^s \left(\mn + \frac{\mq}{x}\right) + 2 \mn \varphi^s_2 $      & $ \varphi_1^s \frac{\sqrt{2}k^R}{x}  $  \\ \hline 
$  \half  $ & $ \varphi_1^s \frac{\sqrt{2}k^L}{x} $ &  $ \varphi_1^s \left(\mn + \frac{\mq}{x}\right) + 2 \mn \varphi^s_2 $  \\
\end{tabular} \caption{$\phi^{\lambda_{_N}}_{\lambda_q }/\sqrt{x}$ for quark scalar diquark configuration. } \label{tab:qswf}
\end{ruledtabular}
\end{table} 
 
  \begin{table}
 \begin{ruledtabular}
\begin{tabular}{ccc}
 &\multicolumn{2}{c}{$\lambda_{_N}$} \\
$\lambda_q, \lambda_a$ & $-\half$ &  $ \half $ \\ \hline
$ -\half,-1  $ & $ -\varphi_1^a \frac{2 k^R}{x(1-x)}  $      & $  0  $   \\ \hline 
$ -\half,0  $ &  $ \varphi_1^a \frac{2\ma}{1-x}-\varphi_2^a\frac{\ma}{\mn}(\mn-\frac{\mq}{x}) $      & $-\varphi_2^a \frac{\ma}{\mn} \frac{\sqrt{2}k^R}{x} $   \\ \hline
$ -\half,+1  $ &  $ -\varphi_1^a \frac{2 k^L}{1-x}  $      & $ \varphi_1^a \sqrt{2} (\mn +\frac{\mq}{x})  $   \\ \hline
$ \half, -1  $ &  $ -\varphi_1^a \sqrt{2} (\mn +\frac{\mq}{x}) $      & $ \varphi_1^a \frac{2 k^R}{1-x} $   \\ \hline 
$  \half,0  $ &  $ \varphi_2^a \frac{\ma}{\mn} \frac{\sqrt{2}k^L}{x}$      & $ - \varphi_1^a \frac{2\ma}{1-x} + \varphi_2^a\frac{\ma}{\mn}(\mn-\frac{\mq}{x})  $   \\ \hline
$  \half,+1  $ &  $0 $      & $ \varphi_1^a \frac{2 k^L}{x(1-x)}  $   \\ \hline
\end{tabular} \caption{$\phi^{\lambda_{_N}}_{\lambda_q \lambda_a}/\sqrt{x}$ for quark axial diquark configuration. } \label{tab:qawf}
\end{ruledtabular}
\end{table}

\begin{widetext}
\begin{eqnarray}
f_{1s} &=& \int \frac{d x d \vec{k}_\perp}{16\pi^3 x^2(1-x)} \left\{\varphi_{1}^{s'} \varphi_{1}^{s} \left[\left(\mq + x\mn\right)^2+\kperp{k}^2-\frac{(1-x)^2}{4}Q^2 \right] 
+ \left(\varphi_{1}^{s'} \varphi_{2}^{s} +\varphi_{2}^{s'} \varphi_{1}^{s} \right)2 x \mn   \left(\mq + x \mn \right)  + \varphi_{2}^{s'} \varphi_{2}^{s} 4 x ^2 \mn^2   \right\}  \notag \\ 
\\ 
f_{2s} &=& \int \frac{\mn d x d \vec{k}_\perp}{8\pi^3 x^2(1-x)} \left\{\varphi_{1}^{s'} \varphi_{1}^{s}  \left(\mq + x\mn\right)(1-x)  
+ \left(\varphi_{1}^{s'} \varphi_{2}^{s} +\varphi_{2}^{s'} \varphi_{1}^{s} \right) x (1-x) \mn  +\left(\varphi_{1}^{s'} \varphi_{2}^{s} -\varphi_{2}^{s'} \varphi_{1}^{s} \right) 2 x \mn \frac{\kperp{k}\cdot\kperp{q}}{Q^2}   \right\}   \\ 
f_{As} &=& \int \frac{d x d \vec{k}_\perp}{16\pi^3 x^2(1-x)} \Big\{\varphi_{1}^{s'} \varphi_{1}^{s} \left[\left(\mq + x\mn\right)^2-\kperp{k}^2+\frac{(1-x)^2}{4}Q^2 \right]  \\
\quad\quad && + \left(\varphi_{1}^{s'} \varphi_{2}^{s} +\varphi_{2}^{s'} \varphi_{1}^{s} \right)2 x \mn   \left(\mq + x \mn \right)  + \varphi_{2}^{s'} \varphi_{2}^{s} 4 x ^2 \mn^2   \Big\}  \notag \\ 
f_{1a} &=& \int \frac{d x d \vec{k}_\perp}{16\pi^3 x^2(1-x)} \bigg\{ 2 \varphi_{1}^{a'} \varphi_{1}^{a}\, \left[  \left(\kperp{k}^2-\frac{(1-x)^2}{4}Q^2\right)\frac{1+x^2}{(1-x)^2} +\frac{2x^2 \ma^2}{(1-x)^2} +\left(\mq + x\mn\right)^2  \right] \\ 
\quad\quad && + \left(\varphi_{1}^{a'} \varphi_{2}^{a} +\varphi_{2}^{a'} \varphi_{1}^{a} \right) \frac{2 x \ma^2}{(1-x)\mn} \left(\mq - x \mn \right)  + \varphi_{2}^{a'} \varphi_{2}^{a} \frac{\ma^2}{ \mn^2}   \left[\left(\mq - x\mn\right)^2+\kperp{k}^2-\frac{(1-x)^2}{4}Q^2 \right]  \bigg\}  \notag \\ 
f_{2a} &=& \int \frac{(-)\mn d x d \vec{k}_\perp}{4\pi^3 x^2(1-x)} \left\{\varphi_{1}^{a'} \varphi_{1}^{a} x \left(\mq + x\mn\right)   
+ \left(\varphi_{1}^{a'} \varphi_{2}^{a} +\varphi_{2}^{a'} \varphi_{1}^{a} \right) \frac{x\ma^2}{\mn}  + \varphi_{2}^{a'} \varphi_{2}^{a} \frac{\ma^2}{2\mn^2} (1-x)\left(\mq - x\mn\right)       \right\}   \\ 
f_{Aa} &=& \int \frac{d x d \vec{k}_\perp}{16\pi^3 x^2(1-x)} \bigg\{2 \varphi_{1}^{a'} \varphi_{1}^{a}\,   \left[ \left(\kperp{k}^2-\frac{(1-x)^2}{4}Q^2\right)\frac{1+x^2}{(1-x)^2} +\frac{2x^2 \ma^2}{(1-x)^2} -\left(\mq + x\mn\right)^2   \right] \\ 
\quad\quad && + \left(\varphi_{1}^{a'} \varphi_{2}^{a} +\varphi_{2}^{a'} \varphi_{1}^{a} \right) \frac{2 x \ma^2}{(1-x)\mn} \left(\mq - x \mn \right)  + \varphi_{2}^{a'} \varphi_{2}^{a} \frac{\ma^2}{ \mn^2}   \left[\left(\mq - x\mn\right)^2 - \kperp{k}^2+\frac{(1-x)^2}{4}Q^2 \right]  \bigg\}  \notag
\end{eqnarray}
Note the expressions in these form factor involving $\varphi_2^a$ are different, as well as the $\varphi_1^{a'} \varphi_1^a$. The former is because we use a different wave function for $\varphi_2^a$, while the latter is because we use $\bar{\varepsilon}$ instead of $\varepsilon$ in defining wave functions involving axial diquark (cf.~discussion in Sec.~\ref{subsec:barequark}). To simplify the presentation, the quark wave functions' dependence on the integration variables are implicit: $\varphi_1^{s}\equiv \varphi_1^s(x,\vec{k}_{i\perp})$ and $\varphi_1^{s'} \equiv \varphi_1^s(x,\vec{k}_{f\perp})$ with $\vec{k}_{f\perp} \equiv \kperp{k}-\frac{1-x}{2} \kperp{q}$ and $\vec{k}_{f\perp} \equiv \kperp{k}+\frac{1-x}{2} \kperp{q}$. Here the integration variable $\kperp{k}$ is shifted from the $\kperp{k}$ in Eqs.~(\ref{eqn:f1s})-(\ref{eqn:fAa}) by $-\frac{1-x}{2}\kperp{q}$. In these expressions, $\mq$, $\ms$, and $\ma$ are the masses of the quark, scalar and axial-vector diquarks. 

\end{widetext}

\section{Hadronic interaction and electroweak current matrix elements}\label{app:hadronicvertices}

The results for $N$-$N$-$\pi$ interaction vertices can be found in Table~\ref{tab:NNpi}. The assignment of intrinsic kinetic variables has been discussed in 
Sec.~\ref{subsec:pioncloudprep}. The metric and Dirac spinor convention used again follows the Lepage-Brodsky in Ref.~\cite{Brodsky:1997de}. 

To calculate matrices elements for $N$-$\Delta$-$\pi$ interaction, we need a convention for spin-$3/2$ Rarita-Schwinger spinors:
\begin{eqnarray}
u_\mu\left(\frac{3}{2}\right) & = &  \varepsilon_\mu (+1) u\left(\half\right) \\ 
u_\mu\left(\frac{1}{2}\right) &= & \sqrt{\frac{1}{3}} \varepsilon_\mu (+1) u\left(-\half\right) + \sqrt{\frac{2}{3}} \varepsilon_\mu (0) u\left(\half\right) \notag \\ 
u_\mu\left(-\frac{1}{2}\right) & = & \sqrt{\frac{2}{3}} \varepsilon_\mu (0) u\left(-\half\right) + \sqrt{\frac{1}{3}} \varepsilon_\mu (-1) u\left(\half\right) \notag \\ 
u_\mu\left(-\frac{3}{2}\right) & = & \varepsilon_\mu (-1) u\left(-\half\right)\ . \notag
\end{eqnarray}
Here the spin projections/helicity projections are labeled as the numbers in parenthesis. Note the vector $\varepsilon_\mu$ is different from the one used in axial-diquark wave function: it satisfies $p_{_\Delta}^\mu \varepsilon_\mu =0 $ \cite{Pasquini:2007iz}. The results are collected in Table~\ref{tab:NDpi}. To compute the current matrix elements, we always choose a frame with $q^+ = 0$. The results in Tables~\ref{tab:NDEMcurrentME},\ref{tab:NDAxialcurrentME}, \ref{tab:DDEMmatrix}, and~\ref{tab:DDAmatrix} are of course boost and rotation invariant (in the transverse plane). 
To reduce space for presentations, only subset of the matrix elements mentioned here are shown, while the others can be inferred using the mirror transformation (w.r.t. to the y-z plane) of these elements ({\it i.e.}, parity transformation multiplied by proper rotation). See the captions of the tables for the details.

  \begin{table}
 \begin{ruledtabular}
\begin{tabular}{ccc}
 &\multicolumn{2}{c}{$\lambda_{_N}$} \\
$\lambda_{_N}'$ & $-\half$ &  $ \half $ \\ \hline
$ -\frac{1}{2}  $ &  $ \frac{i \mn (1-x)}{\sqrt{x}} $  &  $-\frac{i \sqrt{2} k^R}{\sqrt{x}} $ \\ \hline 
 $ \frac{1}{2}  $ & $ \frac{i \sqrt{2} k^L}{\sqrt{x}} $ &  $ \frac{i \mn (x-1)}{\sqrt{x}} $ \\ \hline 
\end{tabular} \caption{ $ \mathcal{V}_{\lambda_{_{Nf}},\lambda_{_{Ni}}} $ using helicity basis.  Note $ \mathcal{V}_{-\lambda_{_{Nf}},-\lambda_{_{Ni}}}( x,k^x,k^y ) = -\mathcal{V}_{\lambda_{_{Nf}},\lambda_{_{Ni}}}( x,-k^x,k^y ) $. Changing the sign of $k^x$ leads to $k^L \leftrightarrow k^R$. This property can be used to infer the matrix elements with positive $\lambda_{_{Ni}}$ based on given matrix elements with negative $\lambda_{_{Ni}}$. 
 } \label{tab:NNpi}
\end{ruledtabular}
\end{table}

  \begin{table}
 \begin{ruledtabular}
\begin{tabular}{cc}
 &\multicolumn{1}{c}{$\lambda_{_N}$} \\
$\lambda_{_\Delta}$ & $-\half$ \\ \hline
$ -\frac{3}{2}  $ &   $  i x^{-3/2} k^R (\md+\mn x)   $ \\ \hline 
$ -\frac{1}{2}  $ & $ -\frac{i x^{-3/2} }{\sqrt{6} \md }\big[(\mn x\!-\!\md) (\md+\mn x)^2\! -\!2 k^L k^R (2 \md\!+\!\mn x)\big] $  \\ \hline 
 $ \frac{1}{2}  $ & $ \frac{i k^L x^{-3/2} }{\sqrt{3} \md } \left[2 k^L k^R-(\mn x-2 \md) (\md+\mn x)\right]  $ \\ \hline 
$ \frac{3}{2}  $ &  $i \sqrt{2} x^{3/2} \left(k^L\right)^2 $ \\ \hline 
\end{tabular} \caption{ $  \mathcal{V}_{\lambda_{_{\Delta }},\lambda_{_{N }}}$ using helicity basis. Note $ \mathcal{V}_{-\lambda_{_{\Delta }},-\lambda_{_{N }}}( x,k^x,k^y ) = + \mathcal{V}_{\lambda_{_{\Delta }},\lambda_{_{N }}}( x,-k^x,k^y )$. Changing the sign of $k^x$ leads to $k^L \leftrightarrow k^R$. This property can be used to infer the matrix elements with positive $\lambda_{_{N}}$ based on given matrix elements with negative $\lambda_{_{N}}$.
 } \label{tab:NDpi}
\end{ruledtabular}
\end{table}

  \begin{table}
 \begin{ruledtabular}
\begin{tabular}{cc}
 &\multicolumn{1}{c}{$\lambda_{_N}$} \\
$\lambda_{_\Delta}$ & $-\half$  \\ \hline
$ -\frac{3}{2}  $ & $ -\frac{q^R}{2} \left[ \ged(\mn-\md) +  \gmd (\mn+\md)\right]   $       \\ \hline 
$ -\half   $ &  $ \frac{q^L q^R }{\sqrt{6}\md} \left[(\ged-\gmd)\md +2\gcd (\mn - \md)\right]    $      \\ \hline
$  \half   $ &  $- \frac{ q^L}{2\sqrt{3}\md} \left[ \ged \md(\mn\!-\!\md)\! -\! \gmd \md (\mn\!+\!\md)\! +\! 4 \gcd q^L q^R\right] $     \\ \hline
$ \frac{3}{2}  $ &  $ \frac{\left(q^L\right)^2}{\sqrt{2}} ( \ged +  \gmd )  $        \\ \hline 
\end{tabular} \caption{ $\mathcal{J}^{(0)V}_{\lambda_{_\Delta}, \lambda_{_N}} (q) \equiv \bar{u}_\alpha(p_{_\Delta},\lambda_{_\Delta}) \Gamma^{\alpha +}_{_{\gamma N; \Delta}}( q,p_{_N}; p_{_\Delta})  u(p_{_N}, \lambda_{_N})/(2 p_{_N}^+)$ using helicity basis. Note $\mathcal{J}^{(0)V}_{-\lambda_{_\Delta}, -\lambda_{_N}} (q) (q^x, q^y) = -\mathcal{J}^{(0)V}_{\lambda_{_\Delta}, \lambda_{_N}} (q) (-q^x, q^y) $ . Changing the sign of $q^x$ leads to $q^L \leftrightarrow q^R$. This property can be used to infer the matrix elements with positive $\lambda_{_{N}}$ based on given matrix elements with negative $\lambda_{_{N}}$.
} \label{tab:NDEMcurrentME}
\end{ruledtabular}
\end{table}

  \begin{table}
 \begin{ruledtabular}
\begin{tabular}{cc}
 &\multicolumn{1}{c}{$\lambda_{_N}$} \\
$\lambda_{_\Delta}$ & $-\half$  \\ \hline
$ -\frac{3}{2}  $ &  $ -\frac{q^R}{2 \mn^2}\left[\ca{4} \md +(2 \ca{3}+\ca{4}) \mn\right]  $  \\ \hline 
$ -\frac{1}{2}  $ &  $ \frac{1}{\sqrt{6} \md \mn^2} \left[\ca{5} \mn^2(\mn+ \md )-2 \ca{3} q^L q^R \mn-\ca{4} \md\, q^L
   q^R \right]$  \\ \hline 
 $ \frac{1}{2}  $ & $-\frac{q^L}{2 \sqrt{3} \md \mn^2} \left[\left(\ca{4} \md (\md + \mn )+2
   (\ca{3}-\ca{5}) \mn^2\right) \right] $ \\ \hline 
$ \frac{3}{2}  $ &  $-\frac{\left(q^L\right)^2}{\sqrt{2} \mn^2} \ca{4}  $  \\ \hline 
\end{tabular} \caption{ $\mathcal{J}^{(0)A}_{\lambda_{_\Delta}, \lambda_{_N}} (q) \equiv \bar{u}_\alpha(p_{_\Delta},\lambda_{_\Delta}) \Gamma^{\alpha +}_{_{A N; \Delta}}( q,p_{_N}; p_{_\Delta})  u(p_{_N}, \lambda_{_N})/(2 p_{_N}^+)$ using helicity basis. Note $\mathcal{J}^{(0)A}_{-\lambda_{_\Delta}, -\lambda_{_N}} (q) (q^x, q^y) = \mathcal{J}^{(0)A}_{\lambda_{_\Delta}, \lambda_{_N}} (q) (-q^x, q^y) $ . Changing the sign of $q^x$ leads to $q^L \leftrightarrow q^R$. This property can be used to infer the matrix elements with positive $\lambda_{_{N}}$ based on given matrix elements with negative $\lambda_{_{N}}$
 } \label{tab:NDAxialcurrentME}
\end{ruledtabular}
\end{table}

  \begin{table*}
 \begin{ruledtabular}
\begin{tabular}{ccc}
 &\multicolumn{2}{c}{$\lambda_{_\Delta}$} \\
$\lambda_{_\Delta}'$ & $-\frac{3}{2}$ &  $ -\frac{1}{2} $ \\ \hline
$ -\frac{3}{2}  $ & $  \Fd{1}+\frac{\Fd{3} q^L q^R}{4 \md^2} $ & $\frac{q^R \left(\md^2 (8 \Fd{1}-4 \Fd{2})+(2
   \Fd{3}-\Fd{4}) q^L q^R\right)}{4 \sqrt{6} \md^3} $ \\ \hline 
$ -\frac{1}{2}  $ &  $\frac{q^L \left(\md^2 (8 \Fd{1}-4 \Fd{2})+(2 \Fd{3}-\Fd{4}) q^L q^R\right)}{4 \sqrt{6} \md^3} $ & 
 $  \frac{(8 \Fd{1}-8 \Fd{2}+\Fd{3}) q^L q^R}{12 \md^2}+\Fd{1}+\frac{(\Fd{3}-\Fd{4})
   \left(q^L\right)^2 \left(q^R\right)^2}{6 \md^4} $ \\ \hline 
 $ \frac{1}{2}  $ & $ \frac{\left(q^L\right)^2 \left(\md^2 (\Fd{3}-4 \Fd{2})-\Fd{4} q^L q^R\right)}{4 \sqrt{3} \md^4} $ &
  $ -\frac{q^L \left(8 \md^4 (\Fd{2}-2 \Fd{1})+\md^2 (8 \Fd{2}-4 \Fd{3}+\Fd{4}) q^L q^R+2
   \Fd{4} \left(q^L\right)^2 \left(q^R\right)^2\right)}{12 \sqrt{2} \md^5} $\\ \hline 
$ \frac{3}{2}  $ &  $ -\frac{\Fd{4} \left(q^L\right)^3}{4 \sqrt{2} \md^3} $ & $ \frac{\left(q^L\right)^2 \left(\md^2 (\Fd{3}-4
   \Fd{2})-\Fd{4} q^L q^R\right)}{4 \sqrt{3} \md^4} $ \\ 
\end{tabular} \caption{ $  \mathcal{J}^{(0)\mathrm{EM}}_{\lambda_{_{\Delta }},\lambda_{_{\Delta }}}$ using helicity basis. Note $ \mathcal{J}^{(0)\mathrm{EM}}_{-\lambda_{_{\Delta }}',-\lambda_{_{\Delta }}}( q^x,q^y ) =  \mathcal{J}^{(0)\mathrm{EM}}_{\lambda_{_{\Delta }}', \lambda_{_{\Delta }}}(-q^x, q^y) $. Changing the sign of $q^x$ leads to $q^L \leftrightarrow q^R$. This property can be used to infer the matrix elements with positive $\lambda_{_\Delta}$ based on given matrix elements. } \label{tab:DDEMmatrix}
\end{ruledtabular}
\end{table*}

 \begin{table*}
 \begin{ruledtabular}
\begin{tabular}{ccc}
 &\multicolumn{2}{c}{$\lambda_{_\Delta}$} \\
$\lambda_{_\Delta}'$ & $-\frac{3}{2}$ &  $ -\frac{1}{2} $ \\ \hline
$ -\frac{3}{2}  $ & $ -\FAd{1}-\frac{\FAd{3} q^L q^R}{4 \md^2}  $ &  $-\frac{q^R \left(4 \FAd{1} \md^2+\FAd{3} q^L q^R\right)}{2
   \sqrt{6} \md^3} $ \\ \hline
$ -\frac{1}{2}  $ & $ -\frac{q^L \left(4 \FAd{1} \md^2+\FAd{3} q^L q^R\right)}{2 \sqrt{6} \md^3} $ & $ \frac{1}{12}
   \left(\frac{(\FAd{3}-8 \FAd{1}) q^L q^R}{\md^2}-4 \FAd{1}-\frac{2 \FAd{3} \left(q^L\right)^2
   \left(q^R\right)^2}{\md^4}\right) $ \\ \hline
 $  \frac{1}{2}  $ & $ -\frac{\FAd{3} \left(q^L\right)^2}{4 \sqrt{3} \md^2} $ & $ 0 $ \\ \hline
$ \frac{3}{2}  $ & $ 0 $ & $ \frac{\FAd{3} \left(q^L\right)^2}{4 \sqrt{3} \md^2}$ \\ 
\end{tabular} \caption{ $  \mathcal{J}^{(0)\mathrm{A}}_{\lambda_{_{\Delta }},\lambda_{_{\Delta }}}$ using helicity basis. Note $ \mathcal{J}^{(0)\mathrm{A}} _{-\lambda_{_{\Delta }}',-\lambda_{_{\Delta }}}( q^x,q^y ) = - \mathcal{J}^{(0)A}_{\lambda_{_{\Delta }}', \lambda_{_{\Delta }}}(-q^x, q^y) $. Changing the sign of $q^x$ leads to $q^L \leftrightarrow q^R$. This property can be used to infer the matrix elements with positive $\lambda_{_\Delta}$ based on given matrix elements. } \label{tab:DDAmatrix}
\end{ruledtabular}
\end{table*}

\newpage

\begin{figure*}[htbp]
\includegraphics[width=\textwidth,angle=0]{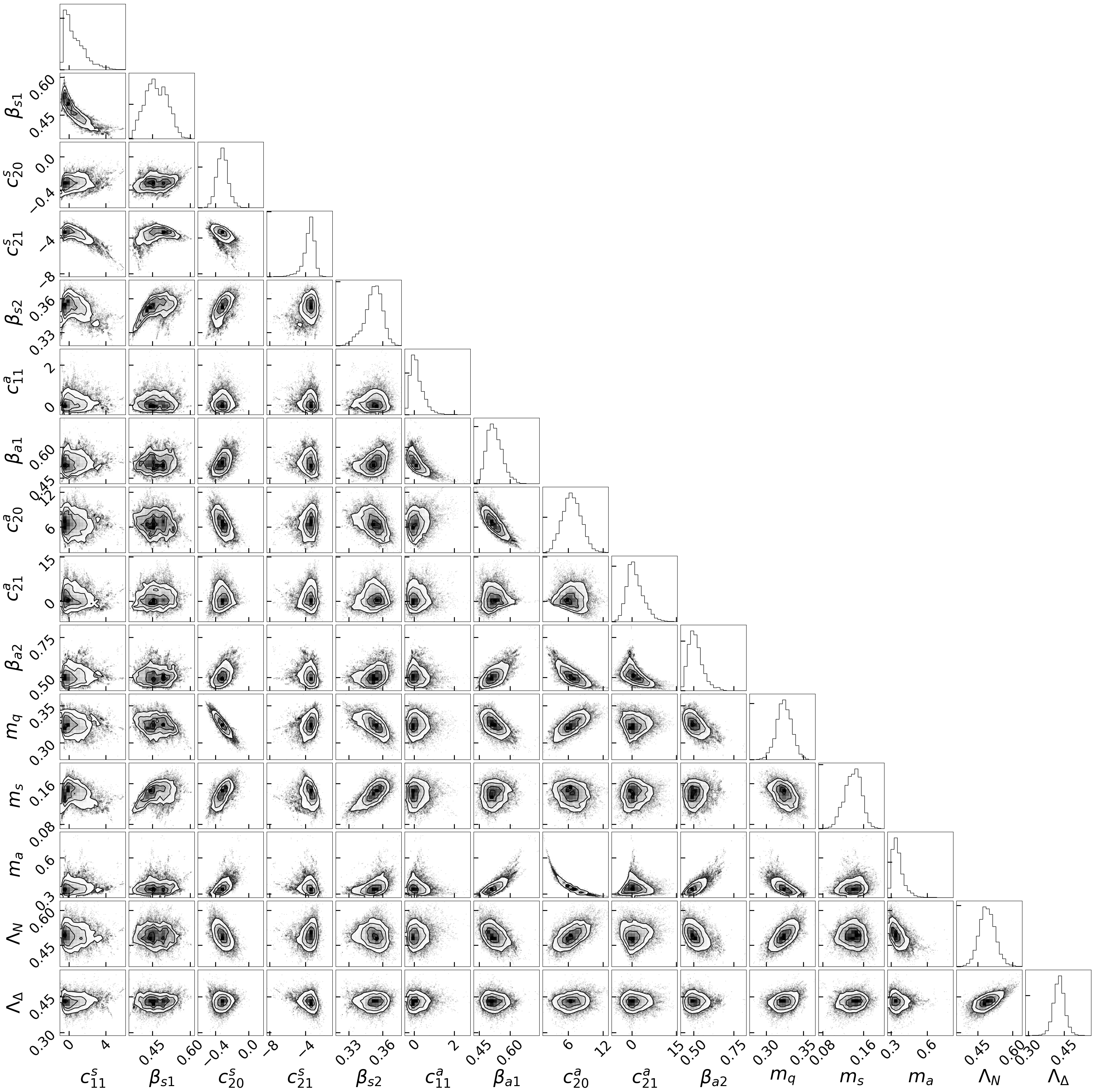} 
\caption{\label{fig:corrplt} The 2-dim and 1-dim projection of the 15-dim PDF, as computed through Bayesian inference. }
\end{figure*}

\clearpage


%\bibliography{../../HadronicPhysics}

\end{document}